\documentclass{article}

\usepackage{arxiv}

\usepackage[utf8]{inputenc} 
\usepackage[T1]{fontenc}    
\usepackage{hyperref}       
\usepackage{url}            
\usepackage{booktabs}       
\usepackage{amsfonts}       
\usepackage{nicefrac}       
\usepackage{microtype}      
\usepackage{lipsum}		
\usepackage{graphicx}
\usepackage{natbib}
\usepackage{doi}
\usepackage{caption}
\usepackage{subcaption}

\usepackage{mathtools,upgreek,eucal,mathrsfs,enumitem,amsthm,amsmath}
\usepackage{times,epsfig,makeidx,amsfonts,amsmath,dcolumn,multirow,amssymb,colortbl,bm,url}
\usepackage{algorithm}
\usepackage[]{algorithmic}
 \usepackage[flushleft]{threeparttable}

\newtheorem{theorem}{Theorem}

\newcommand{\bxi}{\bm{\xi}}

\newcommand{\bpsi}{\bm{\psi}}

\newcommand{\btheta}{\bm{\theta}}
\newcommand{\bdelta}{\bm{\delta}}
\newcommand{\bmeta}{\bm{\eta}}
\newcommand{\bbeta}{\bm{\beta}}
\newcommand{\balpha}{\bm{\alpha}}

\newcommand{\blambda}{\bm{\lambda}}

\newcommand{\bt}{{\bf t}}

\newcommand{\bX}{{\bf X}}
\newcommand{\bx}{{\bf x}}

\title{On near-redundancy and identifiability of parametric hazard regression models under censoring}

\author{
	\href{https://orcid.org/0000-0001-7183-8407}{\includegraphics[scale=0.06]{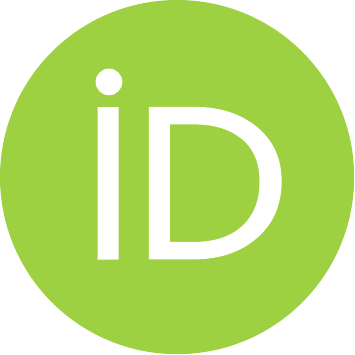}\hspace{1mm}Francisco Javier Rubio} \\
	Department of Statistical Science\\
	University College London \\
	London, UK\\
	\texttt{f.j.rubio@ucl.ac.uk} 
 \And	
	\href{https://orcid.org/0000-0000-0000-0000}{\includegraphics[scale=0.06]{orcid.pdf}\hspace{1mm}Jorge A. Espindola} \\
	Department of Mathematics\\
	University of Sonora \\
	Hermosillo, M{\' e}xico\\
	\texttt{a211220104@alumnos.uson.mx } 
\And
	\href{https://orcid.org/0000-0000-0000-0000}{\includegraphics[scale=0.06]{orcid.pdf}\hspace{1mm}Jos{\'e} A. Montoya} \\
	Department of Mathematics\\
	University of Sonora \\
	Hermosillo, M{\' e}xico\\
	\texttt{montoya@mat.uson.mx} 
	}

\renewcommand{\shorttitle}{XXXX}

\hypersetup{
pdftitle={XXX},
pdfsubject={stat.ME, stat.AP},
pdfauthor={XXX, Rubio},
}

\begin{document}
\maketitle

\begin{abstract}

We study parametric inference on a rich class of hazard regression models in the presence of right-censoring.
Previous literature has reported some inferential challenges, such as multimodal or flat likelihood surfaces, in this class of models for some particular data sets. 
We formalize the study of these inferential problems by linking them to the concepts of near-redundancy and practical non-identifiability of parameters.
We show that the maximum likelihood estimators of the parameters in this class of models are consistent and asymptotically normal. Thus, the inferential problems in this class of models are related to the finite-sample scenario, where it is difficult to distinguish between the fitted model and a nested non-identifiable (\textit{i.e.}, parameter-redundant) model.
We propose a method for detecting near-redundancy, based on distances between probability distributions. We also employ methods used in other areas for detecting practical non-identifiability and near-redundancy, including the inspection of the profile likelihood function and the Hessian method.
For cases where inferential problems are detected, we discuss alternatives such as using model selection tools to identify simpler models that do not exhibit these inferential problems, increasing the sample size, or extending the follow-up time.
We illustrate the performance of the proposed methods through a simulation study. 
\color{black} Our simulation study reveals a link between the presence of near-redundancy and practical non-identifiability. \color{black} 
Two illustrative applications using real data, with and without inferential problems, are presented.
\end{abstract}


\section{Introduction}\label{sec:intro}

The analysis of time-to-event data is of interest in many areas, including medicine, biology, and engineering. Survival data are typically right-censored, as not all of the times-to-event of interest are observed within the follow-up time (administrative censoring) or there may be loss of follow-up of some individuals (random censoring).
In addition to the times-to-event, individual characteristics (covariates) $\bx \in {\mathbb R}^p$ are often available. Thus, it is of interest to incorporate the information in the covariates for modelling the times-to-event. With this aim, a number of survival regression models have been proposed in the literature. Such models are commonly formulated on the hazard function, leading to different hazard structures. The Proportional Hazards (PH) model represents the most popular hazard structure in practice \citep{cox:1972}. 
\color{black} This model is formulated at the hazard level $h(t;\bx, \bbeta)$, and assumes that the covariates $\bx$ have a multiplicative effect on the baseline hazard function $h_0(\cdot)$, as follows
\begin{equation*}
h(t;\bx, \bbeta) = h_0(t)\exp\left\{\bx^{\top}\bbeta\right\},    
\end{equation*}
where $\beta\in{\mathbb R}^p$ are the regression coefficients. \color{black}
Another popular survival regression model is the Accelerated Failure Time (AFT) model, which can be formulated at the hazard level assuming that the covariates have a simultaneous multiplicative effect on the time scale and the hazard function \citep{kalbfleisch:2011}
\begin{equation*}
h(t;\bx, \bbeta) = h_0\left(t \exp\left\{\bx^{\top}\bbeta\right\} \right)\exp\left\{\bx^{\top}\bbeta\right\}.    
\end{equation*}
The AFT model can also be formulated as a log-linear regression model. A richer hazard structure, referred to as the General Hazard (GH) hereafter,  that contains the PH and AFT models, as particular cases, was proposed independently by \cite{etezadi:1987} and \cite{chen:2001}. 
\color{black} The GH model is also formulated at the hazard level $h(t;\bx, \balpha, \bbeta)$, and incorporates multiplicative effects on the time scale and the hazard scale:
\begin{equation}
h(t;\bx, \balpha, \bbeta) = h_0\left(t \exp\left\{\bx^{\top}\balpha\right\} \right)\exp\left\{\bx^{\top}\bbeta\right\},
\label{eq:GHO}
\end{equation}
where $\balpha\in{\mathbb R}^p$ and $\beta\in{\mathbb R}^p$ are the regression coefficients, and $h_0(\cdot)$ is the baseline hazard. \color{black} 
This sort of models have been used in survival analysis applications, as they represent a simple alternative for including time-level effects (through the coefficients $\balpha$), while allowing for a tractable estimation.  These applications include the analysis of the survival of cancer patients \citep{li:2015,rubio:2019B,rubio:2022,amaral:2023}, recurrent event data \citep{xu:2020}, joint models for longitudinal and survival data \citep{rubio:2019S,alvares:2021}, reliability analysis \citep{gamiz:2011}, and as a mean for comparing competing nested sub-models (\textit{e.g.}~PH and AFT) of interest \citep{zhao:2009}. 

Inference on the parameters of the GH model can be conducted using parametric \citep{rubio:2019S} and semiparametric approaches \citep{chen:2001}. 
The parametric approach represents an attractive option as parametric models are interpretable and relatively easy to implement.
However, since parametric models impose specific assumptions on the shape of the baseline hazard $h_0(t;\bxi)$, it is desirable to use flexible models that can capture a variety of shapes of the hazard function to avoid imposing restrictive assumptions. Flexible parametric models require the inclusion of additional parameters. This inclusion of parameters typically carries a cost in the finite-sample inference. For instance, by increasing the uncertainty on the model parameters (wider confidence intervals or wider profile likelihoods) \citep{raue:2009}. For GH models, some references have reported the presence of flat likelihoods or a ``challenging'' optimization of the likelihood function \citep{rubio:2019S,burke:2020}, including non-convergence of some numerical optimization methods or singular Hessian matrices, when the sample size is small or moderate, or with high censoring rate.  
We focus on the study of these inferential problems from a more formal perspective, linking them to the concepts of ``near-redundancy of parameters'' and ``practical non-identifiability'' \citep{catchpole:1997,catchpole:1998,raue:2009,cole:2020}. 

Near-redundancy and practical non-identifiability refer to cases where the statistical inference may fail due to the inability to estimate the parameters of a model. This may be due to data quality, which makes the problem sample-dependent rather than a general shortcoming of a parametric model. A near-redundant model is a model that, formally (\textit{i.e.}~theoretically), is not parameter-redundant, but might behave as a  parameter-redundant model, because it is very similar to a model that is parameter-redundant for a particular data set \citep{cole:2020}. This occurs when the parameter estimates are close to a redundant nested model \citep{cole2012determining} (that is, a model that can be written in terms of a smaller set of parameters). 
On the other hand, a similar and related problem is that of practical non-identifiability. A model is practically non-identifiable if the log-likelihood has a unique maximum, but the length of the individual parameter's likelihood-based confidence region tends to infinity in either or both directions \citep{raue:2009}. These concepts have been largely used in Statistical Ecology and Biology \citep{gimenez:2003,cole:2012,cole:2016,simpson:2022}.
Although these two definitions are presented as separate concepts, we will show that there is an intersection between them in the context of the GH model \eqref{eq:GHO}.

The inferential challenges reported in the use of GH models, discussed above, can be formally classified as problems with near-redundancy of parameters and practical non-identifiability. 
Therefore, it is of interest to develop tools to identify these types of inferential problems in the GH model. 
Methods for detecting near-redundancy or practical non-identifiability are classified into symbolic or numerical methods \citep{catchpole:1997,cole:2020}. The symbolic method or symbolic differentiation method, is a method based on symbolic algebra that can be used to determining whether or not a model is parameter-redundant or non-identifiable \citep{catchpole:2001}. The symbolic method basically consists in determining the rank of matrix of derivatives of exhaustive summary terms \citep{cole:2020}. This method has been used for different classes of models. However, for structurally complex models, computer algebra packages may not be able to calculate the required symbolic rank of the derivative matrices, due to computer memory limitations \citep{cole2010determining,cole2012determining, cai2021parameter}. In such cases, numerical methods have been used as an alternative. These numerical methods include the Hessian method, the simulation method, data cloning, and the profile log-likelihood method (see \citealp{cole:2020} for a discussion on these methods).

In this article we focus on identifying inferential problems associated with the parameter estimation in GH models in the presence of right-censored observations. 
For GH models, we are able to pinpoint the nested non-identifiable model (\textit{i.e.}, parameter-redundant).
Consequently, we can link the inferential problems with the closeness of the estimated model to the nested non-identifiable model. More specifically, we propose a general method for detecting near redundancy based on measuring the distance between the fitted model and the nested non-identifiable model. We develop two criteria, based on the Kullback-Leibler divergence and the Hellinger distance, which take into account the sample size and censoring rate.
In section \ref{sec:hazreg}, we present the parametric general hazard regression model and discuss the identifiability of parameters in this model. Section \ref{sec:inference} presents a result on the consistency and asymptotic normality of the maximum likelihood estimators of the parameters in the GH regression model, and describes some methods used to detect near-redundancy or practical non-identifiability. This section also presents our proposed methodology to detect near-redundancy, which is based on using the Hellinger distance and/or the Kullback-Leibler divergence. In section \ref{sec:simulation}, we conduct a simulation study to investigate the effect of the sample size and the censoring rate on the inference on the parameters of the GH regression model, as well as the performance of the proposed methodology for detecting inferential problems. In section \ref{sec:applications}, we apply the proposed methodology for detecting near-redundancy and practical non-identifiability of parameters in two real-data examples. Finally, we discuss the results obtained in this work in section \ref{sec:discussion}.

\section{Parametric General Hazard Regression Models}\label{sec:hazreg}

In this section, we describe the GH model and discuss several choices for the baseline hazard. We also discuss identifiability of parameters in this model, which we will later connect with the concepts of near-redundancy and practical non-identifiability.

Let us first introduce some notation. Let $o_i \in \mathbb{R}_+$ be the survival times (or times-to-event) for individuals $i=1,\ldots,n$. Let $c_i \in \mathbb{R}_+$ be right-censoring times, such that we only observe $o_i \leq c_i$, $\delta_i=\mbox{I}(o_i\leq c_i)$ be the indicator that observation $i$ is uncensored,  $t_i=\min\{o_i,c_i\}$ be the observed times, and $n_o= \sum_{i=1}^n \delta_i$ be the number of uncensored observations. Let $\bx_i= (x_{i1},\ldots,x_{ip})^{\top} \in{\mathbb R}^p$ denote the vectors of available covariates. Consider the parametric general hazard (GH) structure \citep{chen:2001,rubio:2019S}:
\begin{eqnarray}
h(t ; \btheta) = h_0\left( t \exp \left\{ \tilde{\bx}_i^{\top}\balpha \right\}  ; \bxi \right)\exp \left\{ \bx_i^{\top}\bbeta \right\},
\label{eq:GH}
\end{eqnarray}
where $\btheta = (\bxi^{\top},\balpha^{\top},\bbeta^{\top})^{\top}$; $\bx_i \in {\mathbb R}^p$ are the covariates that have an effect at the hazard-level; $\tilde{\bx}_i\in {\mathbb R}^q$ are the covariates that have an effect at the time-level, with $\tilde{\bx}_i\subseteq \bx_i$ typically; $h_0(\cdot ; \bxi)$ is a parametric baseline hazard function, with parameter $\bxi \in \Xi \subseteq {\mathbb R}^r$. 
The hazard structure \eqref{eq:GH} is slightly more general than \eqref{eq:GHO}, as \eqref{eq:GH} allows for the inclusion of different covariates as time-level and hazard-level effects.
The GH structure \eqref{eq:GH} contains the Proportional Hazards (PH, $\balpha=0$), the Accelerated Failure Time (AFT, $\balpha=\bbeta$ and $\tilde{\bx}=\bx$), and the Accelerated Hazards (AH, $\bbeta=0$) models as particular cases (see \citealp{rubio:2019S} for a more extensive discussion on the interpretation of this hazard structure). The corresponding cumulative hazard function can be written as:
\begin{eqnarray}
H(t ; \btheta) = H_0\left( t \exp \left\{ \tilde{\bx}_i^{\top}\balpha \right\}  ; \bxi \right)\exp \left\{ \bx_i^{\top}\bbeta - \tilde{\bx}_i^{\top}\balpha \right\}.
\label{eq:GCH}
\end{eqnarray}
Common (three-parameter) choices for the baseline hazard are: the Power Generalized Weibull (PGW), Exponentiated Weibull (EW), Generalized Gamma (GG), among other distributions \citep{alvares:2021} (see the Supplementary Material for more information on these distributions). These distributions can capture the basic shapes of the hazard function: increasing, decreasing, unimodal, and bathtub. Simpler (two-parameter) models such as the gamma, log-normal, and log-logistic distributions can also be employed, although they impose more restrictions on the shape of the baseline hazard. These models are implemented in the R package `HazReg` (https://github.com/FJRubio67/HazReg) for several choices of the parametric baseline hazard.

\subsection*{Identifiability}
Recall that a model is identifiable if two different sets of parameter values do not result in the same model \citep{lehmann:2006,cole:2020}. More formally, let ${\mathcal {P}}=\{P_{\theta }:\theta \in \Theta \}$ be a statistical parametric model with parameter space $\Theta$. The model ${\mathcal {P}}$ is said to be identifiable if the mapping $\theta \mapsto P_{\theta }$ is one-to-one \citep{lehmann:2006}. That is, $P_{\theta _{1}}=P_{\theta _{2}}$ implies that $\theta _{1}=\theta _{2}$ for all $\theta _{1},\theta _{2}\in \Theta$. 
The model defined by the GH structure \eqref{eq:GH} is identifiable except for the case when the baseline hazard corresponds to the Weibull distribution \citep{chen:2001}, \color{black}provided that there is no collinearity between the covariates and the baseline hazard is identifiable\color{black}. Indeed, for the Weibull baseline hazard the PH, AFT, and AH models coincide \citep{chen:2001}, making (some or all of) the parameters $\balpha$ and $\bbeta$ redundant. 
Consequently, in the context of the GH model \eqref{eq:GH}, the concepts of parameter-redundancy and non-identifiability are equivalent.
\color{black} To obtain a theoretically identifiable GH model, it is required to use an identifiable baseline hazard which does not belong to the Weibull family, and that there is no collinearity between the covariates\color{black}. However, some flexible parametric distributions such as the PGW, EW and GG distributions contain the Weibull distribution as a particular case (when the parameter $\gamma = 1$ in these distributions, see Supplementary Material). Consequently, the GH model with baseline hazard given by these distributions contains a nested non-identifiable model, for $\gamma = 1$, while it is identifiable for $\gamma\neq 1$. 
\color{black}General families of distributions containing the Weibull distribution as a particular case (or as a limit case) are discussed in \cite{sinner:2022}. We also refer the reader to \cite{ley:2021} for a general discussion on basic desiderata in the use of flexible parametric models. Indeed, non-identifiability, along with near redundancy and practical non-identifiability, make parametric models fail one of their key requirements, namely, ``Straightforward parameter estimation''. \color{black}

\section{Inference in the General Hazard model}\label{sec:inference}

In this section, we present the likelihood function associated with the GH model \eqref{eq:GH} and discuss maximum likelihood estimation (MLE) of the parameters of this model. We present a result on the consistency and asymptotic normality of the MLEs, under standard regularity conditions, which shows that the GH model has good asymptotic properties. Then, we present several methods for detecting practical non-identifiability and near-redundancy of parameters in this model. 
\subsection{Maximum likelihood estimation}
The availability of the hazard and cumulative hazards \eqref{eq:GH}-\eqref{eq:GCH} allows for a tractable implementation of the log-likelihood function, 
\begin{eqnarray*}
\ell( \btheta; \bX, \bt, \bdelta) = \sum_{i=1}^n \delta_i \log h(t_i ; \btheta) -  \sum_{i=1}^nH(t_i ; \btheta),
\end{eqnarray*}
where $\bX = (\bx_1,\dots,\bx_n)$ is the design matrix, $\bt = (t_1,\dots,t_n)^{\top}$, and $\bdelta = (\delta_1,\dots,\delta_n)^{\top}$. Thus, parameter estimation can be performed by using numerical methods, provided the baseline hazard is not Weibull, to guarantee identifiability of parameters. Although this is not an onerous condition, in practice, inferential problems with these models have been reported for small and moderate samples or high censoring rates \citep{rubio:2019S,burke:2020}, even when using baseline hazards different from the Weibull distribution. 
We will show that these inferential problems are related to flat likelihoods and/or likelihood surfaces with multiple local maxima, which complicate the numerical estimation of the parameters. This allows us to connect these inferential challenges with the concepts of  near-redundancy and practical non-identifiability of parameters \citep{cole:2020}, which are finite-sample problems. In contrast, the following result shows that the maximum likelihood estimators of the parameters in the GH model are consistent and asymptotically normal.
\begin{theorem} \label{th:asymp}
Consider the hazard regression model defined by the hazard structure \eqref{eq:GH} with parametric baseline hazard $h_0\left(\cdot ; \bxi \right)$.
Let $ \btheta^{\star} = ({\bxi^{\star}}^{\top},{\balpha^{\star}}^{\top},{\bbeta^{\star}}^{\top})^{\top}$ be the true values of the parameter.
Under conditions C1--C7 in the Supplementary Material, it follows that 
\begin{itemize}
\item[(i)] Consistency: $\widehat{\btheta} \stackrel{P}{\to} \btheta^{\star}$ as $n\to \infty$.
\item[(ii)] Asymptotic normality: $\sqrt{n}\left(\widehat{\btheta} -  \btheta^{\star} \right) \stackrel{d}{\to} N({\bf 0},{\bf I}(\btheta^{\star})^{-1})$ as $n\to \infty$, where ${\bf I}(\btheta^{\star})$ is the matrix $  \operatorname{cov} \left\lbrace\nabla_{\btheta} m_1(\btheta) \right\rbrace$, with $m_1(\btheta)= \delta_1 \log h(t_1 ; \btheta) -  H(t_1 ; \btheta)$, evaluated at $\btheta^{\star}$. 
\end{itemize}
\end{theorem}
Conditions C1-C3 are satisfied by models of practical interest, such as the PGW, GW, and EW distributions. The remaining conditions are standard regularity conditions on parametric models.
This theorem indicates that the model defined by the GH structure \eqref{eq:GH} has good asymptotic properties. However, near-redundancy and practical non-identifiability represent finite-sample problems, which is the scenario of interest in real-life problems. We now discuss general and \textit{ad hoc} methods for detecting these inferential problems in the GH model.

\subsection{Detecting Practical Non-Identifiability: the Profile Likelihood}\label{subsec:profile}
As explained in Section \ref{sec:intro}, if a model exhibits practical non-identifiability, the likelihood surface will contain flat, or nearly flat, ridges. Thus, naturally, a method for detecting these inferential problems consists of visualizing or evaluating the profile likelihood function associated with each parameter. This allows for individually identifying which parameters are involved in the practical non-identifiability problem. This method has been used in biological models \citep{raue:2009}, epidemiological models \citep{tonsing2018profile}, or capture-recapture models \citep{lebreton2002multistate}. For completeness, we present the definition, using our notation on hazard regression models, of the profile likelihood below. We refer the reader to \cite{sprott:2008} for an extensive discussion on the use of the relative profile likelihood in statistical inference.

Suppose that the vector of model parameters $\btheta$ can be decomposed into two subsets of parameters $(\bpsi^{\top},\blambda^{\top})^{\top}$, where $\bpsi=(\psi_1,\ldots,\psi_{d_\psi})^{\top}$ denotes the parameters of interest and $\blambda=(\lambda_1,\ldots,\lambda_{d_\lambda})^{\top}$ are the nuisance parameters. Let also $L( \btheta; \bX, \bt, \bdelta) = \exp \left\{\ell( \btheta; \bX, \bt, \bdelta)\right\}$ denote the likelihood function of $\btheta$. The profile likelihood function of $\bpsi$ is
\begin{equation}
L_P (\bpsi; \bX, \bt, \bdelta)=\sup_{\blambda} L(\bpsi,\blambda; \bX, \bt, \bdelta).
\label{eq:profile}
\end{equation}
Now, the relative profile likelihood function of $\bpsi$ is a standardized version of \eqref{eq:profile}, which takes a value of one at the maximum of the profile likelihood function of $\bpsi$,
\begin{equation*}
R_P (\bpsi; \bX, \bt, \bdelta)=\frac{L_P (\bpsi; \bX, \bt, \bdelta)}{\sup_{\bpsi} L_P(\bpsi; \bX, \bt, \bdelta)}.
\end{equation*}

A level $c$ profile likelihood region for $\bpsi$ is given by
\begin{equation*}
 \left\lbrace \bpsi : R_P (\bpsi; \bX, \bt, \bdelta) \geq c\right\rbrace,
\end{equation*}
where $0\leq c\leq 1$. Since $-2 \log R_P (\bpsi; \bX, \bt, \bdelta)$ follows an asymptotically $\chi^2_{d_\psi}$ distribution \citep{sprott:2008}, a profile likelihood region for $\bpsi$ with approximately 95\% confidence is obtained when $c\approx 0.147$.

The profile likelihood method, for identifying practical non-identifiability, consists of checking if the length of the profile likelihood intervals, of a level of interest $c$, for a specific parameter are infinite in either or both directions. 
The profile likelihood function is rarely available in closed-form, but evaluating this function numerically is usually feasible \citep{van:2000}.
The profile likelihood (equivalently, profile log-likelihood) method is a suitable numerical method for detecting practically non-identifiable models, as it allows inspecting each parameter individually and detecting the parameters producing inferential problems. 
We will explore the use of the profile likelihood for detecting practical non-identifiability in our simulation study and real-data applications.

\subsection{Detecting near-redundancy: Distance-based Methods}\label{subsec:distmethods}
Inferential problems may also appear when the true generating model is theoretically identifiable but it is close to a non-identifiable model. In the GH model, the nested non-identifiable model corresponds to the case where the baseline hazard belongs to the Weibull family. Inferential problems associated with this closeness are thus related to the concept of near-redundancy of parameters. Indeed, the PGW, EW, and GG distributions contain the Weibull distribution as a particular case, and consequently are prone to producing near-redundancy. 

A possible way for detecting when the fitted model is close to the nested non-identifiable model consists of measuring the similarity between the fitted baseline hazard and the Weibull family. 
The similarity between the fitted model and the closest Weibull model can be calculated by using a distance or a divergence between probability measures.
\color{black}
Let ${F}_0 = F_{{\bxi}}$ be the parametric cumulative distribution associated with the baseline hazard in model \eqref{eq:GH}.
\color{black}
Let $\widehat{F}_0 = F_{\widehat{\bxi}}$ be the fitted distribution function (associated with the fitted baseline hazard), with corresponding density function $\widehat{f}_0$, and let $F_W$ be the Weibull distribution function with parameters $\bmeta = (\sigma,\nu)^{\top}$, with density function $f_W$. Following this line of thought, let us define the minimum distance between the fitted model and the Weibull family as:
\begin{equation}
{\mathcal D}(\widehat{F}_0,F_W) = \min_{\bmeta \in {\mathbb R}^2_+}  d(\widehat{F}_0,F_W),
\label{eq:distcrit}
\end{equation}
for some distance or divergence between probability measures $d(\cdot,\cdot)$. In order to use this criterion in practice, we need to specify a distance or divergence between distributions, and a threshold to identify when the two distributions are close. 
\color{black} That is, we would like to define a threshold $U(n,c,d)$ (which depends on the sample size $n$, the number of censored observations $c$, and the distance or divergence $d$) such that if the following inequality holds,
\begin{equation*}
{\mathcal D}(\widehat{F}_0,F_W) \leq U(n,c,d),
\end{equation*}
then the sample and the model are classified as near-redundant (subject to stochastic error).

Regarding the choice of $d(\cdot,\cdot)$, there exist a number of distances (or divergences) between distributions that could be employed \citep{gibbs:2002}, but it is also desirable to use an interpretable distance. The threshold to identify problematic cases should also be linked to the sample size since, as the sample size grows, inferential problems also disappear or become less likely, given that the GH model has good asymptotic properties. Note that if the true value of the parameter $\bxi^{\star}$ (see Theorem \ref{th:asymp}) is such that $F_{{\bxi}^{\star}}$ belongs to the Weibull family, then 
\begin{equation*}
{\mathcal D}(\widehat{F}_0,F_W) \stackrel{P}{\to} {\mathcal D}(F_{{\bxi}^{\star}},F_W) =0, \quad \text{as}  \quad n \to \infty.
\end{equation*}
Consequently, a desirable property of the threshold $U(n,c,d)$ is $\lim_{n\to\infty} U(n,c,d) = 0$.
\color{black}

Next, we present some specific choices for the distance $d(\cdot,\cdot)$, and discuss some heuristic approaches to establish a threshold $U(n,c,d)$ to identify near redundant cases. The performance of these criteria will be assessed later in the simulation study.

\begin{itemize}
\item[(a)] The Kullback-Leibler (KL) divergence, where 
\begin{equation*}
 d(\widehat{F}_0,F_W) = d(\widehat{F}_0 \vert \vert F_W) = \int_0^{\infty} \widehat{f}_0(t) \log\dfrac{\widehat{f}_0(t)}{f_W(t)} dt.
\end{equation*}
To establish a threshold to identify near redundant cases, we borrow some concepts from information theory. 

In information theory, a quantity of interest, related to the KL divergence, is the KL minimax redundancy. Under our notation, this is defined as \citep{cover:2006}
$$
R^* = \min_{F_W} \max_{F_{\bxi}} d({F}_{\bxi},F_W),
$$
where $F_{\bxi}$ is the parametric distribution associated with the baseline hazard. \color{black} Several works in information theory have shown that, under regularity conditions, $R^*$ decreases in probability to zero at a rate ${k \log(n)}/{(2 n)}$, where $k$ is the number of parameters (see \citealp{acharya:2012}, and the references therein, for a general review on this kind of results and \cite{barron:1998} for a theoretical treatment of this result). 


Given that the sample may contain censored observations, instead of using the sample size $n$, we will use the effective sample size $n_{e} = n - \rho c$,  where $\rho \in (0,1)$, which accounts for the loss of information due to censoring. 
Then, motivated by the KL minimax redundancy convergence rate, we propose the threshold $U =  M  k \log(n_e)/(2 n_e)$, where $M>0$ is a positive constant specified by the user (arising from the convergence rate of the KL minimax redundancy, which is only defined up to a proportionality constant). Following this line of thought, we propose the following heuristic criterion to detect near-redundancy. We say that, if the following inequality holds,
\begin{equation}
{\mathcal D}(\widehat{F}_0,F_W) \leq \dfrac{M  k \log(n_e)}{2n_e},
\label{Ineq:KLD}
\end{equation}
then the corresponding GH model \eqref{eq:GH} and the data are classified as near-redundant.
In our applications, and based on our simulations, we will focus on the choice $M=0.05$, but other choices are also possible. We also assume that $k$ is the total number of parameters in the GH model (rather than just those in the baseline hazard), as the model parameters are estimated jointly.

\color{black}

\item[(b)] The Hellinger distance, where
$$
 d(\widehat{F}_0,F_W) =\sqrt{\frac{1}{2}\int_0^{\infty}\left(\sqrt{\widehat{f}_0(t)}-\sqrt{f_W(t)}\right)^2dt}.
$$ 

We will now apply some results from \cite{lecam:1973} to establish a criterion to assess the closeness of $\widehat{F}_0$ and the Weibull family based on this distance.
Following the results in \cite{lecam:1973}, we cannot distinguish between two probability distributions $F$ and $G$, without incurring in a maximal error $\kappa \in (0,1)$, if
\begin{equation}
1-  \sqrt{ 1 - (1-d^2(F,G))^{2n} } \geq 2\kappa,
\label{eq:lecam}
\end{equation}
where $n$ is the sample size (see \cite{baraud:2021} for a discussion on this result).
Recall that we are interested in distinguishing between $\widehat{F}_0$ and the closest $F_W$, based on a sample of size $n$ with $n -c$ uncensored observations. \color{black} Replacing the sample size $n$ by the effective sample size $n_e$ and $d$ by ${\mathcal D}$ in Le Cam's criterion \eqref{eq:lecam}, we obtain the following heuristic criterion to detect near-redundancy. We say that, if the following inequality holds
\begin{equation*}
{\mathcal D}(\widehat{F}_0,F_W) \leq \sqrt{1 - \left(4 \kappa - 4 \kappa^2\right)^{\frac{1}{2 n_e}}},
\end{equation*}
then the corresponding GH model \eqref{eq:GH} and the data are classified as near-redundant.
Equivalently, 
\begin{equation}
n_e\leq \frac{\log \left( 1- \left(1-2\kappa\right)^2\right)}{2\log\left(1-{\mathcal D}^2(\widehat{F}_0,F_W)\right)}.
\label{Ineq:Hellinger}
\end{equation}
\color{black}
The parameter $\kappa>0$ controls the maximal error in distinguishing the two probabilities. However, its interpretation as a maximal error is only approximate as the distribution $\widehat{F}_0$ is based on parameter estimates (from a censored sample) and we are considering the (minimum) distance to an entire family of distributions (Weibull).
By fixing $\kappa$ and for a fixed sample size, this inequality provides a heuristic criterion for identifying the closeness of the estimated model to the non-identifiable Weibull case. In our applications and simulations we will explore the use of $\kappa=0.05$.

\end{itemize}
We note that these criteria impose different thresholds than the original criteria as we are using the effective sample size $n_e$.
\color{black} Following the rule in \cite{liu:2012}, the conventional choice for $\rho = 0.5$, which we adopt in both criteria to define the effective sample size. The performance of criteria \eqref{Ineq:KLD} and \eqref{Ineq:Hellinger} under this choice of the values for $M$, $\kappa$, and $\rho$ will be assessed in the simulation study. However, we emphasize that other choices are also possible.  
\color{black}

Note also that the criteria \eqref{Ineq:KLD} and \eqref{Ineq:Hellinger} can be used to either classify near redundant models based on a specific value of $\mathcal{D}$ and $n_e$, or to identify the minimum number of uncensored observations required to reduce problems with near-redundancy based on a specific value of $\mathcal{D}$. Of course, other distances (such as the Total Variation or Wasserstein-$d$ distances) could be used instead. However, this would require proposing appropriate thresholds to identify near-redundancy. 

{One of the limitations of criteria \eqref{Ineq:KLD} and \eqref{Ineq:Hellinger} is that they are based on point estimates of the parameters, and thus are prone to misclassification in cases where the estimates of the parameters exhibit large uncertainty (\textit{e.g.}~for small samples, high censoring rates, or short follow-up). In our applications in Section \ref{sec:applications}, we will also estimate the probability that inequalities \eqref{Ineq:KLD} and \eqref{Ineq:Hellinger} hold by using a nonparametric bootstrap sample of the maximum likelihood estimates (\textit{i.e.}~obtained by resampling with replacement).}

\subsection{Detecting Near-Redundancy: The Hessian Method}

In addition to the distance-based criteria presented in the previous subsection, in our applications and simulations we will consider the Hessian matrix method \citep{gimenez2004methods,little:2010,cole:2020}. This method consists of calculating the ratio of the smallest eigenvalue and the largest eigenvalue of the Fisher Information matrix or the Hessian (of the log-likelihood function) matrix evaluated at the maximum likelihood estimate (see \citealp{cole:2020} for a detailed discussion on this criterion). \color{black} For completeness, we briefly describe the steps in the Hessian method in Algorithm \ref{alg:Hess}, which we quote \textit{verbatim} from \cite{cole:2020}. \color{black}

\begin{algorithm}
\color{black}
\caption{Hessian method \citep{cole:2020}.}
\begin{algorithmic}[1]
\STATE  Find the Hessian matrix using a numerical method.
\STATE  Find the eigenvalues of the Hessian matrix.
\STATE  Standardise the eigenvalues by taking the modulus of the eigenvalues and dividing through by the largest eigenvalue.
\STATE  The model with that data set is near-redundant if the smallest eigenvalue is less than $0.001$.
\end{algorithmic}
\label{alg:Hess}
\end{algorithm}
\color{black}

\color{black}
Several authors have used the threshold $0.001$ to investigate near-redundancy (see \citealp{cole:2020}, chapter 4 for a discussion). This threshold is based on the concept of ``sloppiness'' of a model (see \citealp{cole:2020}, chapter 3 for a discussion on this topic), and the threshold value proposed by \cite{chis:2016} to detect sloppiness. 
Other rules for selecting the threshold are discussed in \cite{cole:2020}, who also suggested that this threshold should depend on the model. 
We will explore the use of the Hessian method in our simulations and applications in sections \ref{sec:simulation} and \ref{sec:applications}.
\color{black}

Although the concepts of near-redundancy of parameters and practical non-identifiability are studied separately, in the context of the GH model there is an overlap between these two definitions. As shown in the following simulation study and applications, when the baseline hazard is sufficiently close to the Weibull family, the model is classified as near-redundant and we generally observe flat and multimodal profile likelihoods. 
Keeping this in mind, the following simulation study illustrates the link between these two definitions.

\section{Simulation study}\label{sec:simulation}

\color{black}
In this section we present a simulation study which aims at evaluating the performance of the distance-based methods proposed in section \ref{subsec:distmethods}, as well as the Hessian method in Algorithm \ref{alg:Hess}, for detecting near-redundancy in the GH model. Practical non-identifiability is diagnosed using the profile likelihood, as discussed in section \ref{subsec:profile}. We evaluate the effect of the sample size and the censoring rate on the presence of near-redundancy and practical non-identifiability. This simulation study also aims at understanding the link between the presence of near-redundancy and practical non-identifiability problems. \color{black} 

\subsection*{Simulation scenarios}

We present seven simulation scenarios in total. In each case, we simulate $M=250$ samples of varying sizes $n= 250, 500, 1000$. For each scenario, we select administrative censoring times that produce censoring rates of approximately $30\%$ and $50\%$. In addition, in all scenarios, we only include one covariate as this is the simplest structure where, if any inferential problems are detected, such problems could only become more severe with the inclusion of more covariates. This covariate is generated from the normal distribution with mean $0$ and standard deviation $1$. We employ the regression parameter values $\alpha=1.5$ and $\beta=2.5$ for all scenarios. 

In the first three scenarios, the times-to-event are simulated from the GH model \eqref{eq:GH} with PGW baseline hazard (denoted as PGW-GH) model using the probability integral transform \citep{rubio:2019S}.
In Scenario 1, we select parameter values that produce a unimodal baseline hazard function $\sigma=0.3$, $\nu=1.5$, $\gamma=5$. This represents a scenario where the baseline hazard shape cannot be captured by the Weibull baseline hazard, and thus no inferential problems would be expected. 
In Scenario 2, we select parameter values that produce an increasing baseline hazard function $\sigma=1.2$, $\nu=1.3$, $\gamma=0.85$. This hazard shape can indeed be obtained with a Weibull hazard, although the hazard tails between the PGW and Weibull distributions differ for $\gamma \neq 1$, and thus the model is theoretically identifiable. 
In Scenario 3, we select parameter values that produce a decreasing baseline hazard function $\sigma=0.1$, $\nu=0.9$, $\gamma=4$. This hazard shape can also be captured by the Weibull baseline hazard.
In Scenario 4,  the data are simulated from a GH model \eqref{eq:GH} with lognormal (LN) baseline hazard (denoted as LN-GH). The shape of the baseline hazard function is unimodal, and we select the parameters $\tilde{\mu} = 0$ and $\tilde{\sigma} = 1.5$ (mean and standard deviation of the distribution on the log scale). This represents a hazard shape that cannot be captured by the Weibull hazard, and the baseline hazard is outside the PGW family.
In Scenario 5, the data are simulated from a GH model \eqref{eq:GH} with Exponentiated Weibull (EW) baseline hazard (denoted as EW-GH). The shape of the baseline hazard function is increasing,  and we select the parameters $\sigma=0.7$, $\nu=1.2$, $\gamma=0.85$. This represents a hazard shape that can be captured by the Weibull hazard, and the baseline hazard is outside the PGW family.
For Scenarios 1--5, we fit the PGW-GH model. Thus, for Scenarios 1--3 we have correct model specification, while for Scenarios 4--5 there is model misspecification. However, the fitted model (PGW-GH) can capture the baseline hazard shapes from the true generating model in these scenarios.  
In Scenario 6, we use the data simulated in Scenario 2 from the PGW-GH model but we fit a GH model \eqref{eq:GH} with Exponentiated Weibull (EW) baseline hazard (denoted as EW-GH). In Scenario 7, we use the data simulated in Scenario 2 from the PGW-GH model but we fit a GH model \eqref{eq:GH} with Generalized Gamma (GG) baseline hazard (denoted as GG-GH). The aim of Scenarios 6--7 is to illustrate that the inferential problems (near-redundancy and practical non-identifiability) are present for models that contain the Weibull baseline hazard, and that these problems are not specific to the PGW-GH model. Figure 1 in the Supplementary Material shows the shape of all the baseline hazards used in this simulation study.

All models are fitted using the R package `HazReg', which is available at \url{https://github.com/FJRubio67/HazReg}. The integral required for criteria \eqref{Ineq:KLD} and \eqref{Ineq:Hellinger} is calculated using the R command \texttt{integrate}. The profile likelihood functions are implemented directly and the optimization step is performed using the R command \texttt{nlminb}. The Hessian matrix, required to implement the Hessian method, is calculated using the R package \texttt{numDeriv}.

\subsection*{Results}
To evaluate the performance of the proposed distance-based criteria \eqref{Ineq:KLD} and \eqref{Ineq:Hellinger}, we calculate the proportion of times that these criteria classify a simulated sample as a sample producing near-redundancy of parameters of the fitted models. For the KL divergence criterion \eqref{Ineq:KLD} we use $M = 0.05$, and for the Hellinger distance criterion \eqref{Ineq:Hellinger} we choose $\kappa = 0.05$.
We have compared other values of $M$, and we found that this value produced good results in terms of accurate classification and indeed producing very similar results to those obtained with criterion \eqref{Ineq:Hellinger}. Thus, a sample is classified as a sample producing near redundancy if the inequalities in criteria \eqref{Ineq:KLD} and/or \eqref{Ineq:Hellinger} hold. \color{black} We also employ the Hessian method in Algorithm \ref{alg:Hess} to identify near-redundancy. \color{black}

\color{black} An important aim of this simulation study is to compare the classification of samples producing near-redundancy against the classification of samples producing practical non-identifiability. \color{black} 
To detect practical non-identifiability, we assess the flatness of the profile likelihoods of $\alpha$ and $\beta$. More specifically, we evaluate the profile likelihoods at $\widehat{\alpha} \pm \Delta_\alpha$ and $\widehat{\beta} \pm \Delta_\beta$, with $\Delta_\alpha = \Delta_\beta = 3$, and verify if these values are below $0.147$ (the level used to construct 95\% profile likelihood confidence intervals). The values of $\Delta_\alpha$ and$ \Delta_\beta$ where chosen by visually inspecting several profile likelihood functions and identifying a suitable conservative length. 

\color{black}
Tables \ref{tab:S1}-\ref{tab:S3} show the 2-way classification of samples producing ``near-redundancy'' (NR) and/or ``practical non-identifiability'' (PNI) problems, using the aforementioned methods. The cases where no NR nor PNI problems are detected, using the aforementioned methods, will be referred to as ``identifiable'' (I).
The reason why results are presented in 2-way tables follows the main aim of the simulation study, which is about connecting and comparing the presence of the two inferential problems of interest (NR and PNI). Note that the true classification of NR and PNI cases is not known beforehand. Thus, the results presented in this section should be read considering the profile method (used to classify PNI problems) as a reference, as this method is based on directly detecting flat ridges in the profile likelihood.
\color{black}

For Scenario 1 (Table \ref{tab:S1}), corresponding to a baseline hazard with unimodal shape that cannot be captured by the Weibull hazard function, we can see that very few samples induce inferential problems, and that there is an interplay between censoring and sample size that produces problematic samples. As expected, the larger the sample or the lower the censoring rate, the fewer problematic samples are observed. 
{For Scenario 2 (Table \ref{tab:S2}), we observe a very large proportion of samples producing near-redundancy and practical non-identifiability, and that the proportion of samples without inferential problems very slowly decreases as the sample size grows or the censoring rate decreases. This is a challenging scenario where it is very difficult to distinguish the fitted model from the nested Weibull (non-identifiable) model. We also notice that there is a non-negligible proportion of cases that are classified as identifiable using criteria \eqref{Ineq:KLD} and/or \eqref{Ineq:Hellinger}. This indicates that the proposed distance-based methods for detecting near redundancy may not necessarily be able to detect practical non-identifiability problems in all samples. That is, these methods are based on point estimates of the parameters, which can be inaccurate (\textit{e.g.}~$\widehat{\gamma}$ far from $1$) for some samples, and provide no indication of near-redundancy. On the other hand, the profile likelihood is able to detect practical non-identifiability problems as it takes the (large) uncertainty in the estimation of the parameters into account. Another interesting behaviour is that $30\%$ censoring rate seems to produce more samples that induce inferential problems than $50\%$ censoring rate, which sounds counter-intuitive. The reason for this is that the estimates of $\gamma$ have much more variability when the censoring rate is $50\%$, so some estimates of this parameter are further away from $\gamma=1$, but also far way from the true value of the parameter (see Tables 3-4 in the Supplementary Material). This scenario shows that not even samples of size $n=1000$ are enough to substantially reduce the presence of inferential issues. In this case, fitting a simpler, identifiable, model would be an alternative. For instance, fitting an AFT model with Weibull baseline hazard, and comparing those models using a model selection tool. We will explore this idea in the applications presented in section \ref{sec:applications}.}
For Scenario 3 (Table \ref{tab:S3}), we notice a much lower proportion of problematic samples (compared to Scenario 2), and a clear reduction in the proportion of inferential problems when the sample size grows or the censoring rate decreases. This is in line with the intuition that suggests that the larger the sample, or the lower censoring rate, the more information is contained in the data to estimate the model parameters (and thus being able to distinguish the fitted model from the nested non-identifiable model). Tables 1-6 in the Supplementary Material show summaries of the MLEs for Scenarios 1--3 that complement the information and comments provided here. 
\color{black}
Interestingly, in Scenarios 1--3 the Hessian method produces very similar performance to the distance-based methods. This is both, reassuring about the results obtained with the distance-based methods, and a favorable outcome. Indeed, rather than considering distance-based methods and the Hessian method as competitors, they represent tractable and interpretable tools to identify near-redundancy problems, which may be jointly used.  
\color{black}

Results for Scenario 4 are presented in Tables 7--9 in the Supplementary Material. We observe a similar performance of the distance-based measures and the profile likelihood function as in Scenario 1, despite model misspecification. The reason for this is that the baseline hazard is also unimodal, even though it is not PGW, a hazard shape that can be closely captured by the PGW distribution. Results for Scenario 5 are presented in Tables 10--12 in the Supplementary Material. We observe a similar performance of the distance-based measures and the profile likelihood function as in Scenario 2. Again, the reason for this is that the baseline hazard is also increasing and close to the Weibull distribution, even though it is not PGW. 
Results for Scenario 6 and 7 are presented in Tables 13--15 and 16--18, respectively, in the Supplementary Material. In both scenarios, we observe high proportions of cases with near-redundancy and practical non-identifiability. MLEs for both scenarios exhibit high bias and large variability. These scenarios show that even under model misspecification, if the fitted model has a baseline hazard that cannot be distinguished from the Weibull hazard, it is possible to observe problems of near-redundancy and practical non-identifiability. 

\color{black}
We have found that the proposed distance-based methods and the Hessian method have a comparable performance in classifying samples into NR and I cases. 
Moreover, Tables \ref{tab:S1}--\ref{tab:S3} indicate a considerable agreement between the classification of ``PNI \emph{vs.} NR'' and ``I \emph{vs.} I'' cases. Consequently, our results reveal a link between the presence (or absence) of practical non-identifiability and near-redundancy problems in the GH model.
However, we have also found cases where the true model is very close to the non-identifiable model (Scenario 2), and the profile likelihood contains flat ridges, but the distance-based methods and the Hessian method do not classify such samples as NR. This indicates that the two concepts (PNI and NR) are not equivalent, at least with the detection methods studied here. Indeed, in Scenario 2, the Hessian method leads to a slightly larger overlap in the classification of ``PNI \emph{vs.} NR'' cases than that obtained with distance-based methods.
Apparently, a reason for this difference is that the Hessian method incorporates information about the curvature of the likelihood function (Hessian matrix) at the MLE, while distance-based methods only utilize the parameter values that maximize the likelihood function (\textit{i.e.}~the MLE). Moreover, distance-based methods and the Hessian method do not account for uncertainty on the parameters as they are based on point estimates (which might be biased).
In our applications in section \ref{sec:applications}, we will explore the idea of incorporating uncertainty about the estimation of the parameters using nonparametric bootstrap. Overall, we recommend combining the distance-based methods and the Hessian method with the inspection of the profile likelihood curves to aid the detection of inferential problems.

\color{black}

\begin{table}[ht]
\centering
        \begin{tabular}{ |c|c c|c c|c c| }
           \cline{1-7}
            \textbf{H}  &  PNI & I &  PNI & I &  PNI & I\\
            \hline
    Cens: $50\%$       &  \multicolumn{2}{c|}{$n=250$}  & \multicolumn{2}{c|}{ $n=500$} &  \multicolumn{2}{c|}{$n=1000$}\\
     \hline
            \multicolumn{1}{ |c|  }{NR}  & 0.028  & 0  & 0 & 0 & 0  & 0 \\
           \multicolumn{1}{ |c|  }{I} & 0.024  &  0.948 & 0  & 1  & 0  & 1  \\ 
           \hline
              Cens: $30\%$       &  \multicolumn{2}{c|}{$n=250$}  & \multicolumn{2}{c|}{ $n=500$} &  \multicolumn{2}{c|}{$n=1000$}\\
     \hline
            \multicolumn{1}{ |c|  }{NR}  & 0.004  & 0  & 0 & 0 & 0  & 0 \\
           \multicolumn{1}{ |c|  }{I} &  0 & 0.996  & 0  & 1  & 0  & 1  \\ 
           \hline
        \end{tabular}
        
        \medskip
        \centering
        \begin{tabular}{ |c|c c|c c|c c| }
           \cline{1-7}
            \textbf{KL}  &  PNI & I &  PNI & I &  PNI & I\\
            \hline
     Cens: $50\%$       &  \multicolumn{2}{c|}{$n=250$}  & \multicolumn{2}{c|}{ $n=500$} &  \multicolumn{2}{c|}{$n=1000$}\\
     \hline
            \multicolumn{1}{ |c|  }{NR}  & 0.028  & 0  & 0 & 0 & 0  & 0 \\
           \multicolumn{1}{ |c|  }{I} & 0.024  &  0.948 & 0  & 1  & 0  & 1  \\  
           \hline
               Cens: $30\%$       &  \multicolumn{2}{c|}{$n=250$}  & \multicolumn{2}{c|}{ $n=500$} &  \multicolumn{2}{c|}{$n=1000$}\\
     \hline
            \multicolumn{1}{ |c|  }{NR}  & 0.004  & 0  & 0 & 0 & 0  & 0 \\
           \multicolumn{1}{ |c|  }{I} &  0 & 0.996  & 0  & 1  & 0  & 1  \\
           \hline
        \end{tabular}
        
        \color{black}
               \medskip
        \centering
        \begin{tabular}{ |c|c c|c c|c c| }
           \cline{1-7}
            \textbf{Hessian}  &  PNI & I &  PNI & I &  PNI & I\\
            \hline
     Cens: $50\%$       &  \multicolumn{2}{c|}{$n=250$}  & \multicolumn{2}{c|}{ $n=500$} &  \multicolumn{2}{c|}{$n=1000$}\\
     \hline
            \multicolumn{1}{ |c|  }{NR}  & 0.024  & 0  &  0 & 0 &  0 & 0 \\
           \multicolumn{1}{ |c|  }{I}       &  0.028 &  0.948 &  0 & 1  &  0 & 1  \\  
           \hline
               Cens: $30\%$       &  \multicolumn{2}{c|}{$n=250$}  & \multicolumn{2}{c|}{ $n=500$} &  \multicolumn{2}{c|}{$n=1000$}\\
     \hline
            \multicolumn{1}{ |c|  }{NR}  &  0.004 & 0  & 0 & 0  & 0  & 0 \\
           \multicolumn{1}{ |c|  }{I}       &  0 & 0.996  & 0  & 1  &  0 & 1  \\  
           \hline
           \end{tabular} \color{black}
        \caption{Simulation study: Scenario 1. Classification of samples into Identifiable (I), Practical non-identifiable (PNI), and Near-redundant (NR). \color{black} The rows represent the classification of NR and I cases, and the columns represent the classification of PNI and I cases. \color{black} \textbf{H} indicates the NR classification based on the Hellinger distance, \textbf{KL} denotes the NR classification based on the Kullback-Leibler divergence, and \textbf{Hessian} denotes the NR classification based on the Hessian method.}
         \label{tab:S1}
\end{table}

\begin{table}[ht]
\centering
        \begin{tabular}{ |c|c c|c c|c c| }
           \cline{1-7}
           \textbf{H}  &  PNI & I &  PNI & I &  PNI & I\\
            \hline
    Cens: $50\%$       &  \multicolumn{2}{c|}{$n=250$}  & \multicolumn{2}{c|}{ $n=500$} &  \multicolumn{2}{c|}{$n=1000$}\\
     \hline
            \multicolumn{1}{ |c|  }{NR}  & 0.896  &  0.004 & 0.840 & 0.008 &  0.828 & 0.016 \\
           \multicolumn{1}{ |c|  }{I} & 0.080  & 0.020  &  0.136 &  0.016 & 0.124  & 0.032  \\ 
           \hline
              Cens: $30\%$       &  \multicolumn{2}{c|}{$n=250$}  & \multicolumn{2}{c|}{ $n=500$} &  \multicolumn{2}{c|}{$n=1000$}\\
     \hline
            \multicolumn{1}{ |c|  }{NR}  & 0.976  & 0.016  & 0.944 & 0.020 & 0.924  & 0.012 \\
           \multicolumn{1}{ |c|  }{I} & 0.008  &  0 &  0.008 &  0.028 & 0.028  &  0.036 \\ 
           \hline
        \end{tabular}
        
        \medskip
        \centering
        \begin{tabular}{ |c|c c|c c|c c| }
           \cline{1-7}
            \textbf{KL}  &  PNI & I &  PNI & I &  PNI & I\\
            \hline
     Cens: $50\%$       &  \multicolumn{2}{c|}{$n=250$}  & \multicolumn{2}{c|}{ $n=500$} &  \multicolumn{2}{c|}{$n=1000$}\\
     \hline
            \multicolumn{1}{ |c|  }{NR}  & 0.892  &  0.008 & 0.848 & 0.008 &  0.844 & 0.020 \\
           \multicolumn{1}{ |c|  }{I} & 0.084  & 0.016  & 0.128  & 0.016  &  0.108 & 0.028  \\ 
           \hline
               Cens: $30\%$       &  \multicolumn{2}{c|}{$n=250$}  & \multicolumn{2}{c|}{ $n=500$} &  \multicolumn{2}{c|}{$n=1000$}\\
     \hline
            \multicolumn{1}{ |c|  }{NR}  & 0.980  & 0.016  & 0.948 & 0.028 &  0.932 & 0.020 \\
           \multicolumn{1}{ |c|  }{I} &  0.004 & 0  & 0.044  & 0.020  & 0.020  &  0.028 \\ 
           \hline
        \end{tabular}

        \color{black}
                        \medskip
        \centering
        \begin{tabular}{ |c|c c|c c|c c| }
           \cline{1-7}
            \textbf{Hessian}  &  PNI & I &  PNI & I &  PNI & I\\
            \hline
     Cens: $50\%$       &  \multicolumn{2}{c|}{$n=250$}  & \multicolumn{2}{c|}{ $n=500$} &  \multicolumn{2}{c|}{$n=1000$}\\
     \hline
            \multicolumn{1}{ |c|  }{NR}  & 0.912  & 0.008  & 0.920 & 0.024 & 0.928  & 0.044 \\
           \multicolumn{1}{ |c|  }{I}    & 0.068  & 0.012  & 0.056  &  0 &  0.028 &  0 \\  
           \hline
               Cens: $30\%$       &  \multicolumn{2}{c|}{$n=250$}  & \multicolumn{2}{c|}{ $n=500$} &  \multicolumn{2}{c|}{$n=1000$}\\
     \hline
            \multicolumn{1}{ |c|  }{NR}  &  0.936 &  0.016 & 0.924 & 0.040 & 0.928  & 0.044 \\
           \multicolumn{1}{ |c|  }{I}    & 0.048  &  0 &  0.028 &  0.008 & 0.024  &  0.004 \\  
           \hline
           \end{tabular}    \color{black} 
\caption{Simulation study: Scenario 2. Classification of samples into Identifiable (I), Practical non-identifiable (PNI), and Near-redundant (NR). \color{black} The rows represent the classification of NR and I cases, and the columns represent the classification of PNI and I cases. \color{black} \textbf{H} indicates the NR classification based on the Hellinger distance, \textbf{KL} denotes the NR classification based on the Kullback-Leibler divergence, and \textbf{Hessian} denotes the NR classification based on the Hessian method.}
         \label{tab:S2}
\end{table}

\begin{table}[ht]
\centering
        \begin{tabular}{ |c|c c|c c|c c| }
           \cline{1-7}
            \textbf{H}  &  PNI & I &  PNI & I &  PNI & I\\
            \hline
    Cens: $50\%$       &  \multicolumn{2}{c|}{$n=250$}  & \multicolumn{2}{c|}{ $n=500$} &  \multicolumn{2}{c|}{$n=1000$}\\
     \hline
            \multicolumn{1}{ |c|  }{NR}  & 0.028  &  0 & 0.04 & 0 & 0  & 0 \\
           \multicolumn{1}{ |c|  }{I} &  0.248 & 0.724  & 0.044  & 0.952  & 0.012  & 0.988  \\ 
           \hline
              Cens: $30\%$       &  \multicolumn{2}{c|}{$n=250$}  & \multicolumn{2}{c|}{ $n=500$} &  \multicolumn{2}{c|}{$n=1000$}\\
     \hline
            \multicolumn{1}{ |c|  }{NR}  &  0.016 &  0 & 0.004 & 0 & 0  & 0 \\
           \multicolumn{1}{ |c|  }{I} & 0.028  & 0.956  & 0.028  & 0.968  & 0.024  &  0.976 \\ 
           \hline
        \end{tabular}
        
        \medskip
        \centering
        \begin{tabular}{ |c|c c|c c|c c| }
           \cline{1-7}
            \textbf{KL}  &  PNI & I &  PNI & I &  PNI & I\\
            \hline
     Cens: $50\%$       &  \multicolumn{2}{c|}{$n=250$}  & \multicolumn{2}{c|}{ $n=500$} &  \multicolumn{2}{c|}{$n=1000$}\\
     \hline
            \multicolumn{1}{ |c|  }{NR}  & 0.028  &  0 & 0.04 & 0 & 0  & 0 \\
           \multicolumn{1}{ |c|  }{I} &  0.248 & 0.724  & 0.044  & 0.952  & 0.012  & 0.988  \\  
           \hline
               Cens: $30\%$       &  \multicolumn{2}{c|}{$n=250$}  & \multicolumn{2}{c|}{ $n=500$} &  \multicolumn{2}{c|}{$n=1000$}\\
     \hline
            \multicolumn{1}{ |c|  }{NR}  & 0.016  &  0 & 0 & 0 &  0 & 0 \\
           \multicolumn{1}{ |c|  }{I} & 0.028  & 0.956  &  0.032 &  0.968 & 0.024  & 0.976  \\ 
           \hline
        \end{tabular}

        \color{black}
                        \medskip
        \centering
        \begin{tabular}{ |c|c c|c c|c c| }
           \cline{1-7}
            \textbf{Hessian}  &  PNI & I &  PNI & I &  PNI & I\\
            \hline
     Cens: $50\%$       &  \multicolumn{2}{c|}{$n=250$}  & \multicolumn{2}{c|}{ $n=500$} &  \multicolumn{2}{c|}{$n=1000$}\\
     \hline
            \multicolumn{1}{ |c|  }{NR}  &  0.068 &  0 & 0.028 & 0.028 & 0.004  & 0.012 \\
           \multicolumn{1}{ |c|  }{I}    & 0.208  & 0.724  & 0.020  & 0.924  &  0.008 & 0.976  \\  
           \hline
               Cens: $30\%$       &  \multicolumn{2}{c|}{$n=250$}  & \multicolumn{2}{c|}{ $n=500$} &  \multicolumn{2}{c|}{$n=1000$}\\
     \hline
            \multicolumn{1}{ |c|  }{NR}  & 0.028  &  0.008 & 0.032 & 0.004 & 0.024  & 0.020 \\
           \multicolumn{1}{ |c|  }{I}    & 0.016  & 0.948  & 0  & 0.964  & 0  & 0.956  \\  
           \hline
           \end{tabular}     \color{black}
\caption{Simulation study: Scenario 3. Two-way classification of samples into Identifiable (I), Practical non-identifiable (PNI), and Near-redundant (NR). \color{black} The rows represent the classification of NR and I cases, and the columns represent the classification of PNI and I cases. \color{black} \textbf{H} indicates the NR classification based on the Hellinger distance, \textbf{KL} denotes the NR classification based on the Kullback-Leibler divergence, and \textbf{Hessian} denotes the NR classification based on the Hessian method.}       
         \label{tab:S3}
\end{table}

\pagebreak

\section{Applications}\label{sec:applications}

This section presents two real-data examples that illustrate the use of the proposed methodology for detecting near-redundancy and practical non-identifiability of parameters. We apply the distance-based methods and the Hessian method presented in section \ref{sec:inference}, as well as the evaluation of the profile likelihood function. 
In the first example, we illustrate a case where the GH model exhibits near-redundancy and practical non-identifiability of parameters due to the closeness of the baseline hazard to the Weibull family. In the second example we present a case without near-redundancy and where the fitted model can be easily distinguished from the Weibull family. The code and data for these examples are available at \url{https://github.com/FJRubio67/NRPNISurv}

\subsection{Case study I: Lung cancer data}
We analyze survival data of patients with advanced lung cancer from the North Central Cancer Treatment Group. The data set was obtained from the \texttt{survival} R package, and contains information about $n=227$ patients. For each patient, the following variables were recorded: survival time in days (converted to years for our analysis), vital status, age in years (standardized for our analysis), gender and ECOG performance score. The 25\%, 50\% and 75\% quantiles of the patients' survival time were $0.456$, $0.699$, and $1.084$ years. Among the patients, $63$ had an ECOG performance score of $0$, $113$ had an ECOG performance score of $1$, $50$ had an ECOG performance score of $2$, and $1$ had an ECOG performance score of $3$. Female patients corresponded to $39\%$ of the total sample.

\begin{figure}[ht]
\centering
\begin{tabular}{c c c}
    \includegraphics[scale = 0.45]{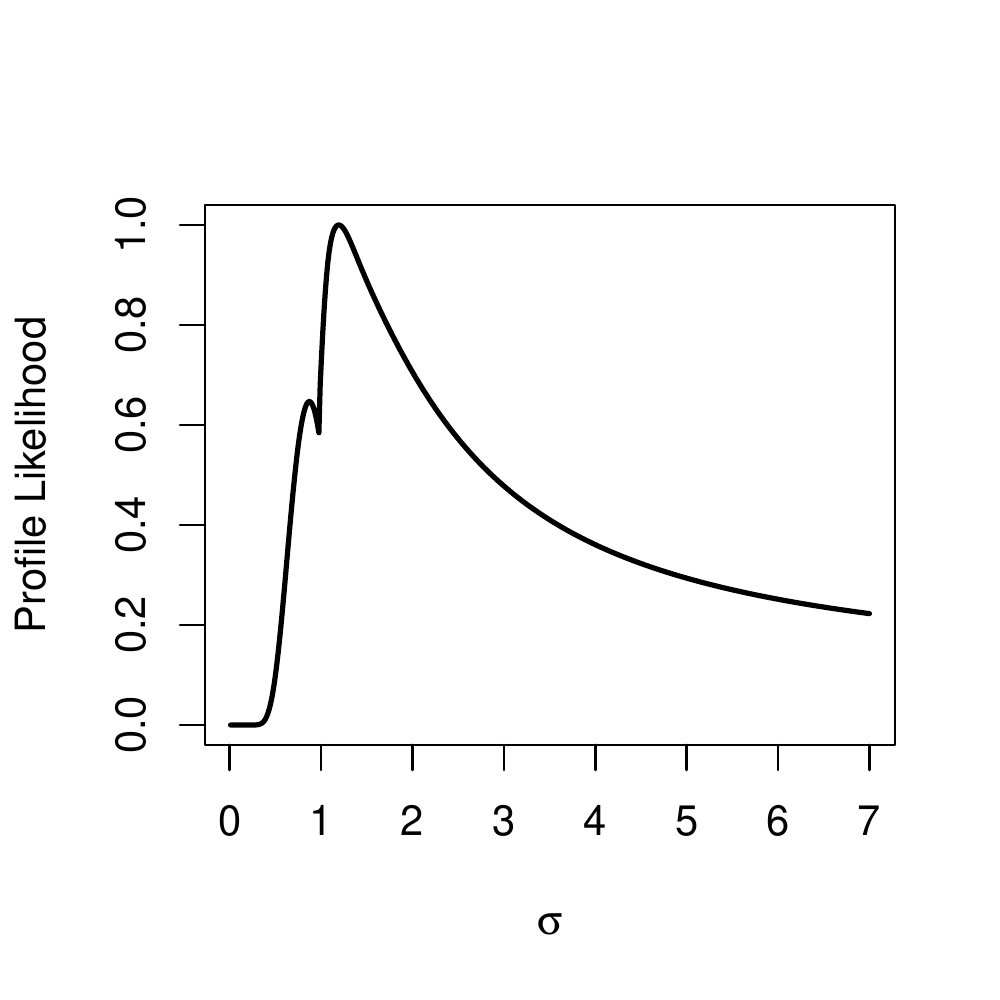} &
    \includegraphics[scale = 0.45]{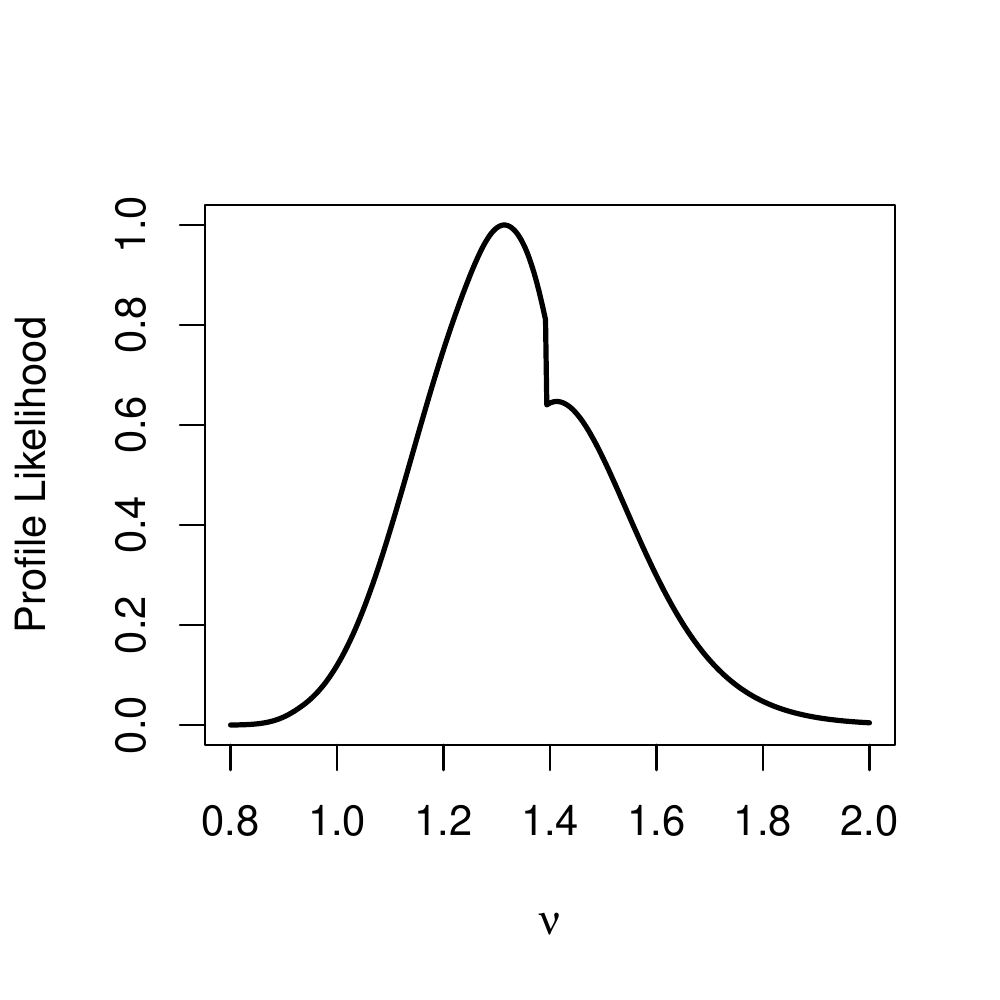} &
    \includegraphics[scale = 0.45]{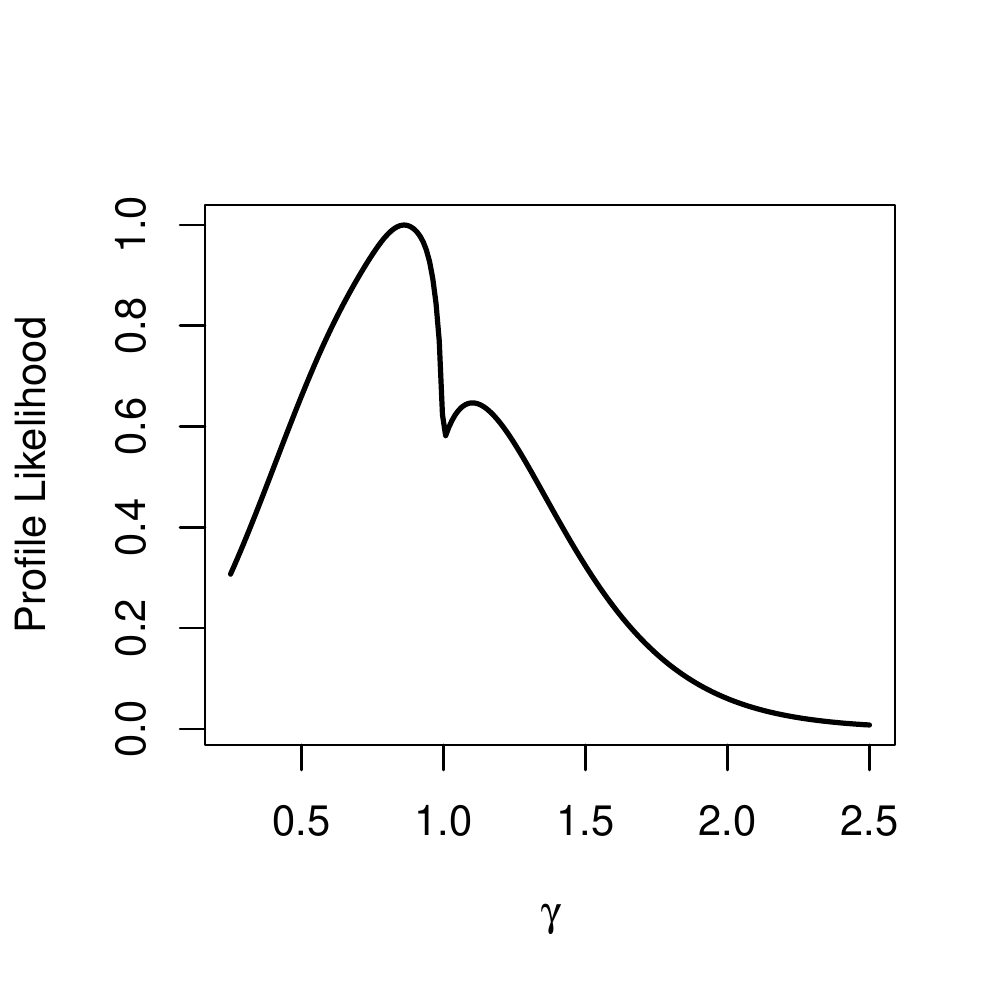} \\
    (a) & (b) & (c) \\
    \includegraphics[scale = 0.45]{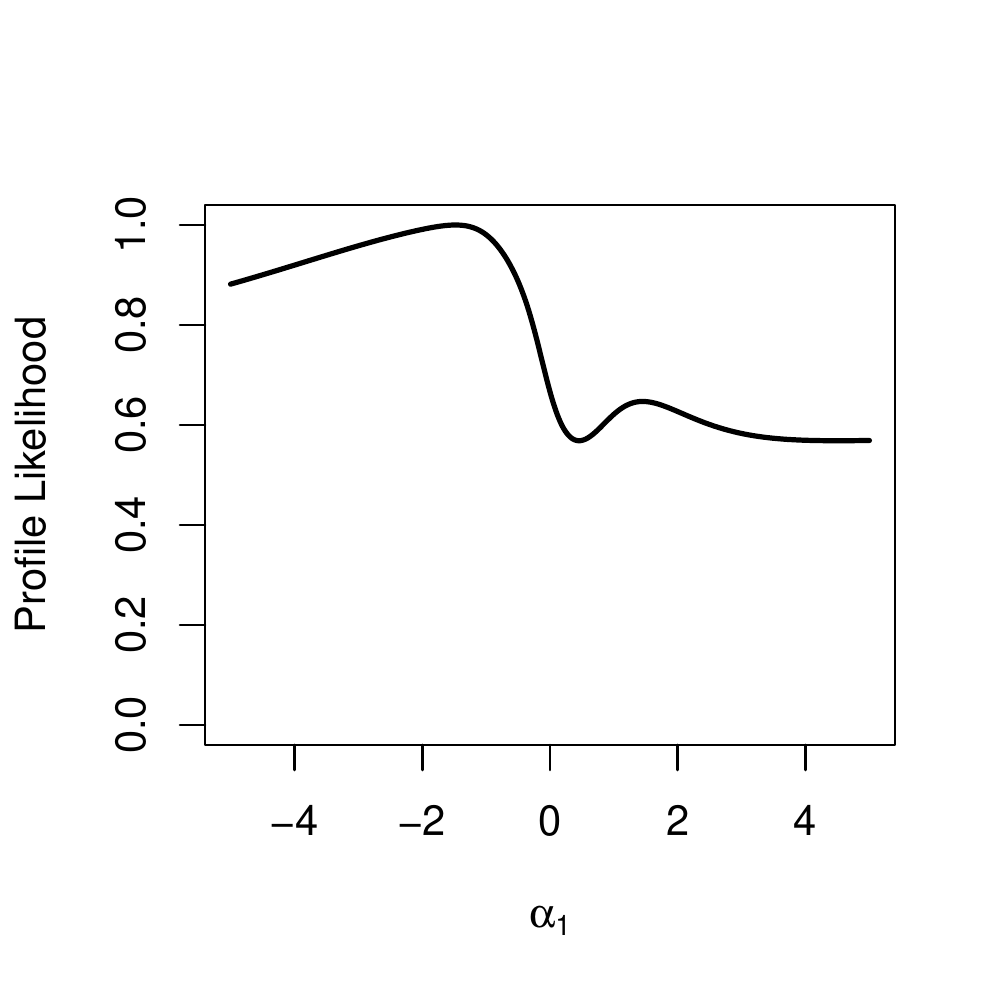} &
    \includegraphics[scale = 0.45]{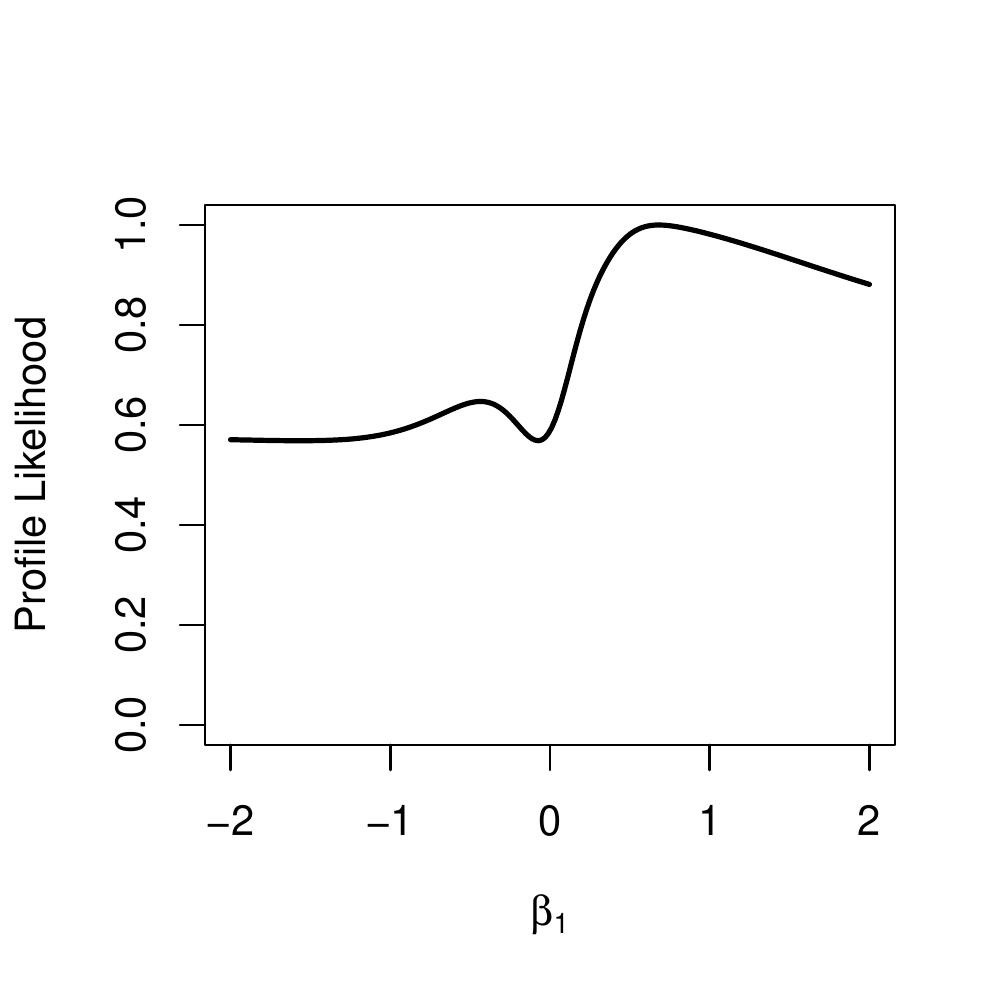} &
    \includegraphics[scale = 0.45]{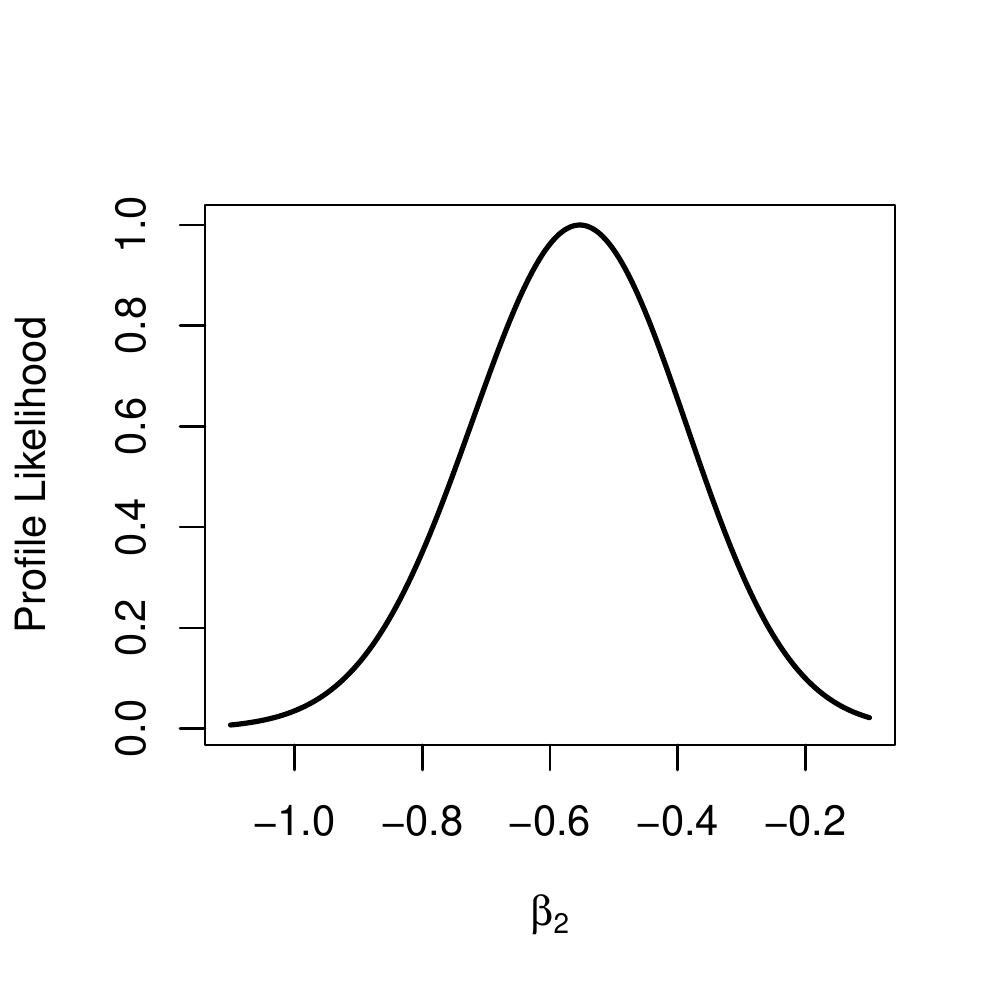} \\
    (d) & (e) & (f) \\
    \includegraphics[scale = 0.45]{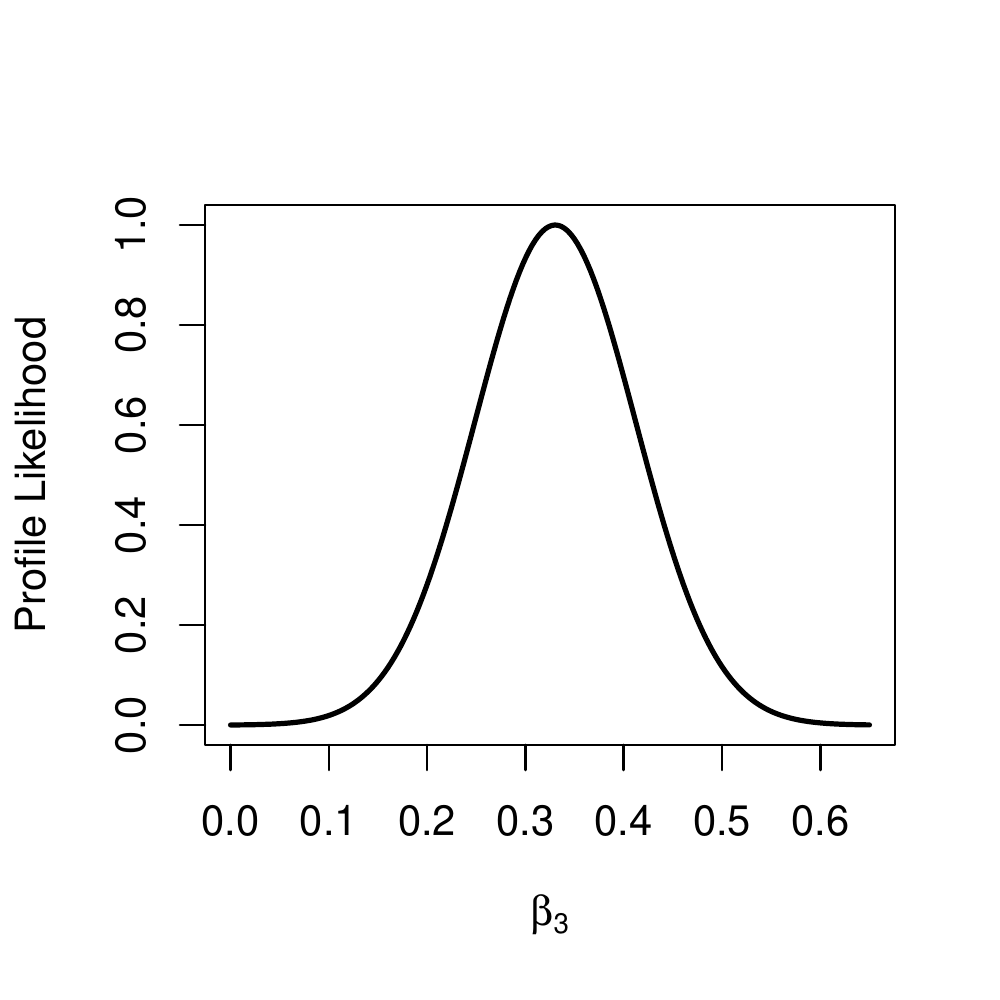} &
     &
     \\
    (g) &  &    
    \end{tabular}
    \caption{Lung cancer data. Profile likelihood plots for the parameters of the PGW-GH model.}
    \label{fig:PGW-GH}
\end{figure}

To analyze this data set, we fit the GH model \eqref{eq:GH} with PGW baseline hazard (denoted as PGW-GH) with time-level covariate \texttt{age}, and hazard-level covariates \texttt{age}, \texttt{sex} and \texttt{ph.ecog} (ECOG performance score). The maximum likelihood estimates of the parameters of this model are shown in the Table \ref{tab:MLE}. In this case $\widehat{\gamma} = 0.861$, which indicates that the fitted model is not a (theoretically) parameter-redundant model (\textit{i.e.}~$\widehat{\gamma}\neq 1$) at this parameter value. The relative profile likelihoods of the parameters are presented in Figure \ref{fig:PGW-GH}. In these plots we can notice that the maximum likelihood estimates exist, in the sense that an overall maximum of the likelihood function is attained. However, we can also notice a number of issues with the profile likelihoods. Regarding the estimation of the parameters of the baseline hazard, the profile likelihood of $\sigma$ has a second mode at a value $\sigma < 1$, and an inflection point around $\sigma \approx 1$. The profile likelihood of $\nu$ is unimodal, but it contains a point at $\nu \approx 1$ where the derivative changes abruptly. From the profile likelihood of $\gamma$, we notice the presence of local maxima, together with an inflection point at $\gamma=1$. We emphasize that these problems are not related to numerical issues \citep{cole2019parameter}, but to changes in the direction of the profile likelihood at specific values of the parameters associated with the nested non-identifiable model ($\gamma=1$). 
For the parameters $\alpha_1$ and $\beta_1$, the relative profile likelihood is flat in both directions, indicating problems of practical non-identifiability of these parameters. The parameters $\alpha_1$ and $\beta_1$ are indeed associated with the same covariate (\texttt{age}). Therefore, even though the maximum likelihood estimator of $\gamma$ is not exactly one, flat and multimodal relative profile likelihoods are obtained for the parameters associated with this covariate. In contrast, the profile likelihoods for the parameters $\beta_2$ and $\beta_3$ are unimodal and concentrated around their maximum.

We now look at the proposed distance-based criteria. The minimum KL divergence between the fitted PGW baseline hazard and the Weibull family is equal to $0.00036$, while the upper bound, with $M=0.05$, in criterion \eqref{Ineq:KLD} is $0.033$, and the effective sample size in $195.5$. Thus, the KL divergence criterion \eqref{Ineq:KLD} suggests near-redundancy of parameters of the PGW-GH model.
Also, the minimum Hellinger distance between the fitted PGW baseline hazard and the Weibull family is equal to $0.0096$. Thus, the Hellinger distance criterion \eqref{Ineq:Hellinger}, with $\kappa=0.05$, also indicates near-redundancy of the parameters of the PGW-GH model. 
{To incorporate the uncertainty in the estimation of the parameters into the distance based criteria \eqref{Ineq:KLD} and \eqref{Ineq:Hellinger}, we apply these criteria to $B= 1000$ bootstrap samples of the MLEs (\textit{i.e.}~obtained via re-sampling with replacement). We estimate the probability of these criteria by taking the proportion of times the inequalities \eqref{Ineq:KLD} and \eqref{Ineq:Hellinger} hold. We found that the KL divergence criterion \eqref{Ineq:KLD} indicates near-redundancy with probability $0.997$, while the Hellinger criterion \eqref{Ineq:Hellinger} indicates near-redundancy with probability $0.92$. Thus, there is a high probability that \eqref{Ineq:KLD} and \eqref{Ineq:Hellinger} hold, suggesting the presence of near-redundancy of parameters.}
On the other hand, the smallest standardized eigenvalue of the Hessian Matrix of the log-likelihood function is $6.5 \times 10^{-5}$.
\color{black} Consequently, the Hessian method (with $0.001$ threshold) indicates near-redundancy of parameters of this model and data. The bootstrap probability of the Hessian method with $0.001$ threshold is $1$, providing further evidence of near-redundancy.\color{black}

To complete our analysis, we consider alternative models: the GH model \eqref{eq:GH} with exponentiated Weibull (EW) baseline hazard (denoted as EW-GH), which represents a competitor of the PGW-GH model in terms of hazard structure and flexibility, a PH model with PGW baseline hazard (PGW-PH), an AFT model with PGW baseline hazard (PGW-AFT), and an AFT model with Weibull baseline hazard (W-AFT).
The EW distribution, like the PGW, is equal to a Weibull distribution if the shape parameter $\gamma$ is equal to one (see Supplementary Material). The maximum likelihood estimates of the parameters of these models are also shown in Table \ref{tab:MLE}. We notice that the MLE of $\gamma$ in the EW model and the PGW model are all different from one. From Table \ref{tab:MLE}, the W-AFT model is favored by the Akaike information criterion (AIC), followed by the PGW-PH model, showing that indeed simpler models are favored by the data.
Although AIC is not a method to detect near-redundancy, we can see that model selection techniques can also be useful to identify simpler models without these inferential problems.

\begin{table}[ht]
\begin{center}
\begin{tabular}{| c | c | c | c | c | c |  }
\hline
 & \multicolumn{5}{ c| }{Model} \\ \hline
Parameter & PGW-GH & PGW-PH & PGW-AFT & EW-GH &  W-AFT\\ \hline
Scale $\widehat{\sigma}$ & 1.194 & 1.310 & 0.803 & 1.162 & 0.984 \\ 
Shape $\widehat{\nu}$ & 1.314 & 1.286 & 1.439   & 1.601 & 1.368 \\ 
Shape $\widehat{\gamma}$ & 0.861 & 0.769 & 1.210   & 0.783 & -- \\ 
$\texttt{age}_t$  $\widehat{\alpha}_1$ & -1.479 & -- & --  & -0.797 & -- \\ 
\texttt{age} $\widehat{\beta}_1$ & 0.681 & 0.093 & 0.070  & 0.439 &  0.068 \\
\texttt{sex} $\widehat{\beta}_2$ & -0.553 & -0.547 & -0.415   & -0.538 & -0.401 \\
\texttt{ph.ecog} $\widehat{\beta}_3$ & 0.330 & 0.335 & 0.251  & 0.326 & 0.243\\  
AIC &  342.357 & 341.169 & 341.250  & 341.644 & \textbf{337.487} \\ \hline
\end{tabular}
\caption{Lung cancer data. Maximum likelihood estimation summary.}
\label{tab:MLE}
\end{center}
\end{table}

\pagebreak
\subsection{Case study II: Leukemia data}
We now analyze the \texttt{LeukSurv} data set from the \texttt{spaBayesSurv} R package. This data set contains information about $n=1043$ patients with acute myeloid leukemia. For each case, the following variables were recorded: survival time in days (converted to years for our analysis), vital status, age in years (standardized for our analysis), gender, white blood cell count (standardized) at diagnosis and Townsend score (standardized). The 25\%, 50\% and 75\% quantiles of the patients' survival time were $0.112$, $0.507$ and $1.468$ years. Female patients corresponded to 48\% of the sample.

\begin{figure}[ht]
\centering
\begin{tabular}{c c c}
    \includegraphics[scale = 0.45]{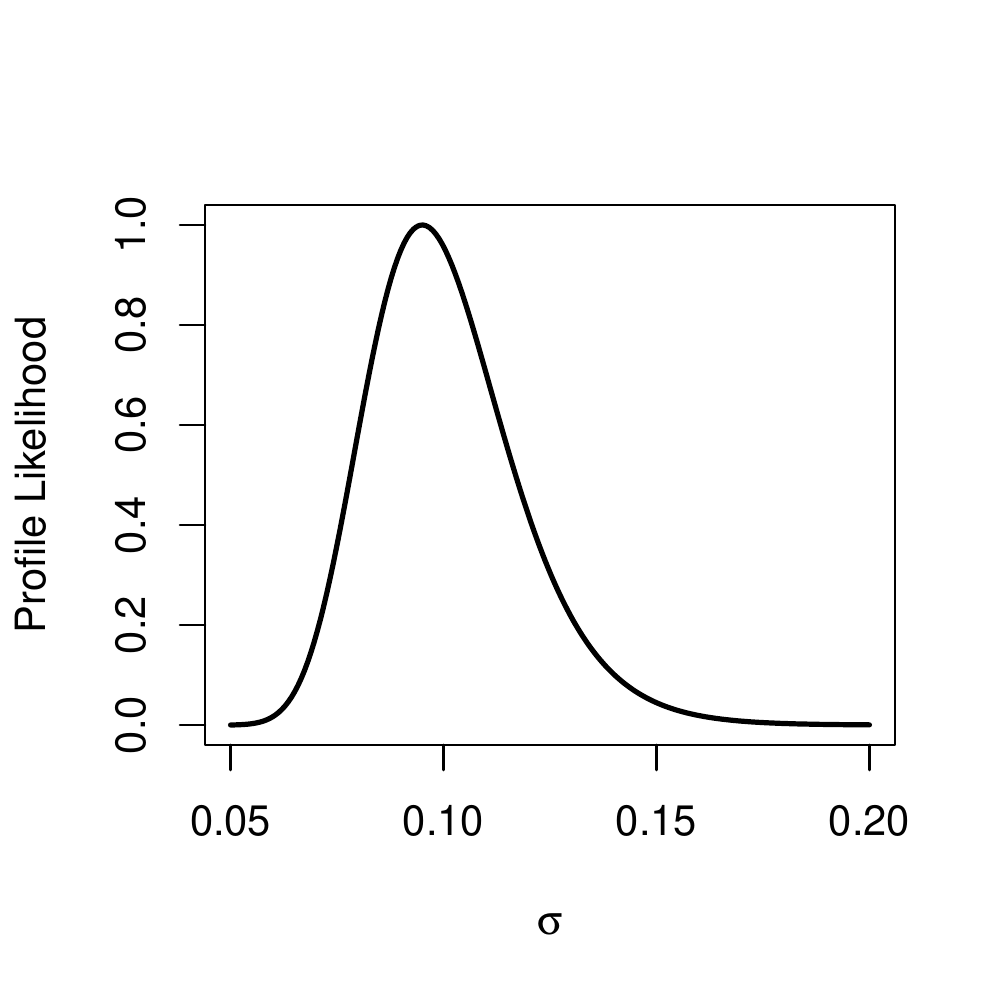} &
    \includegraphics[scale = 0.45]{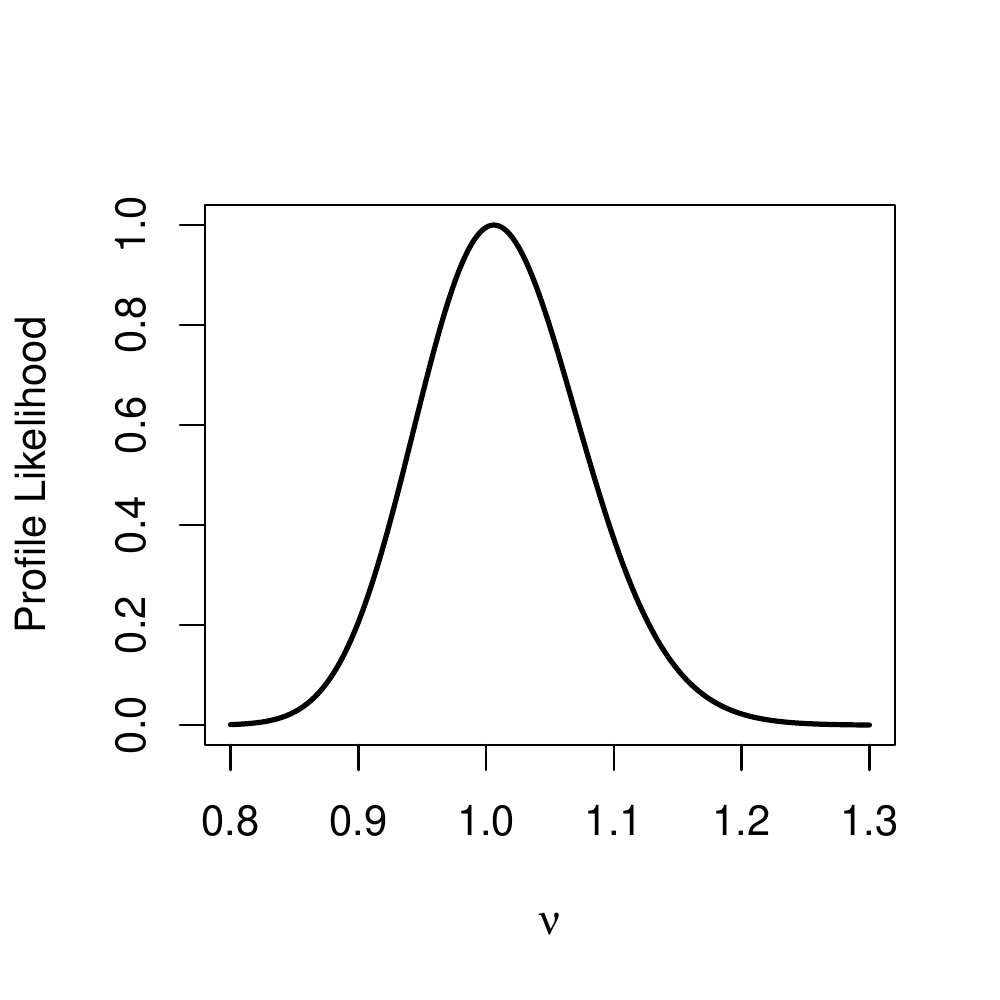} &
    \includegraphics[scale = 0.45]{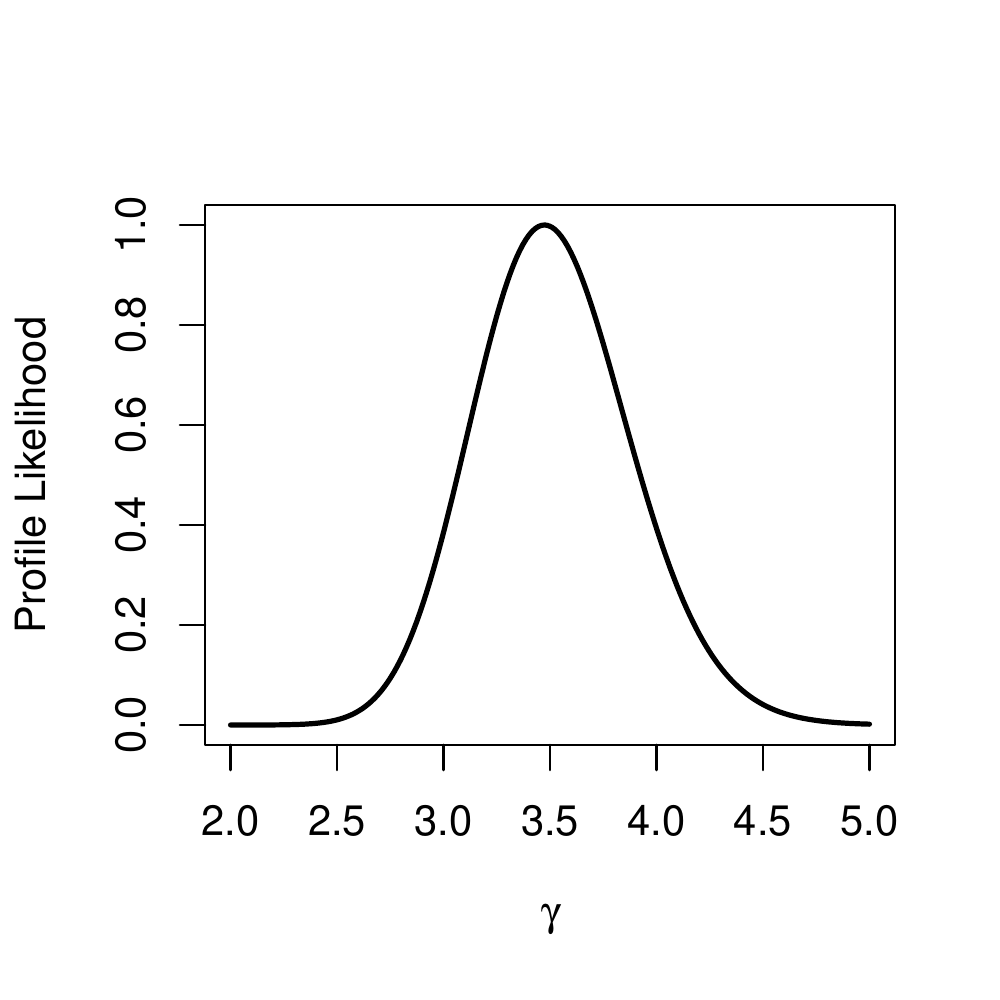} \\
    (a) & (b) & (c) \\
    \includegraphics[scale = 0.45]{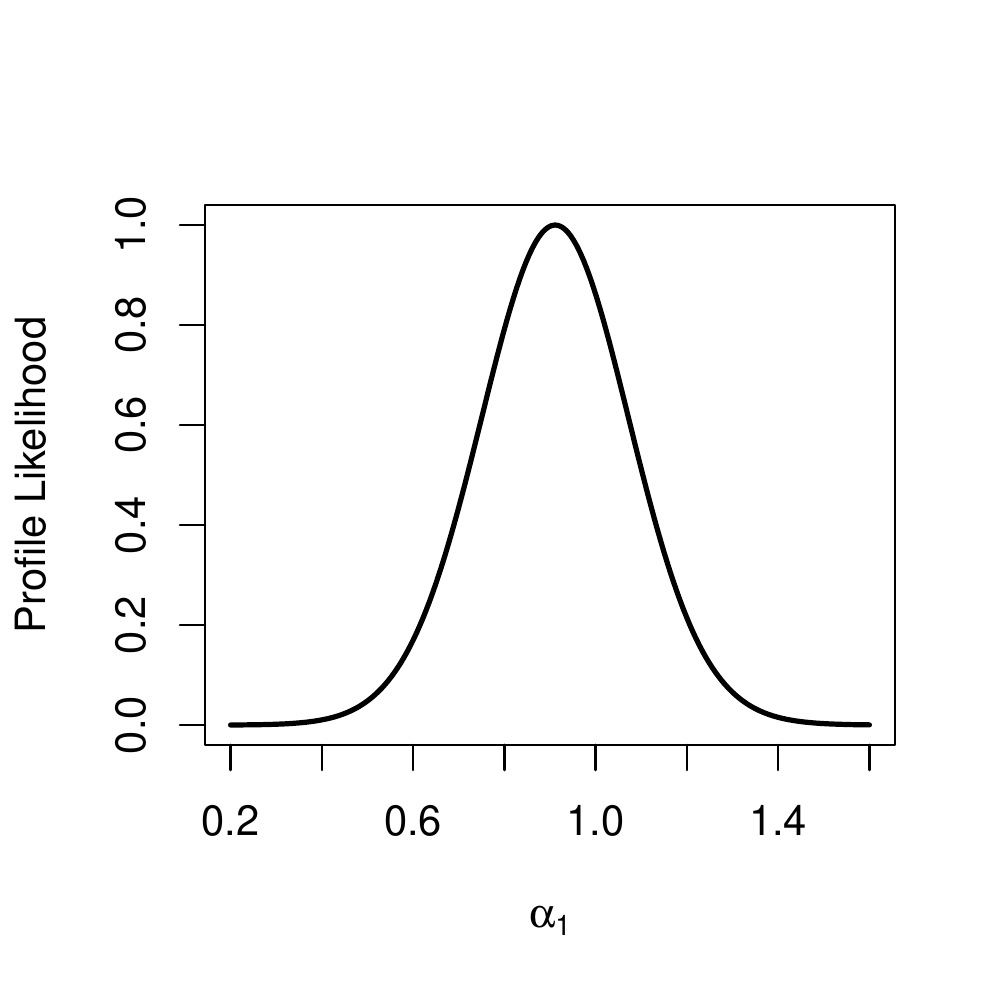} &
    \includegraphics[scale = 0.45]{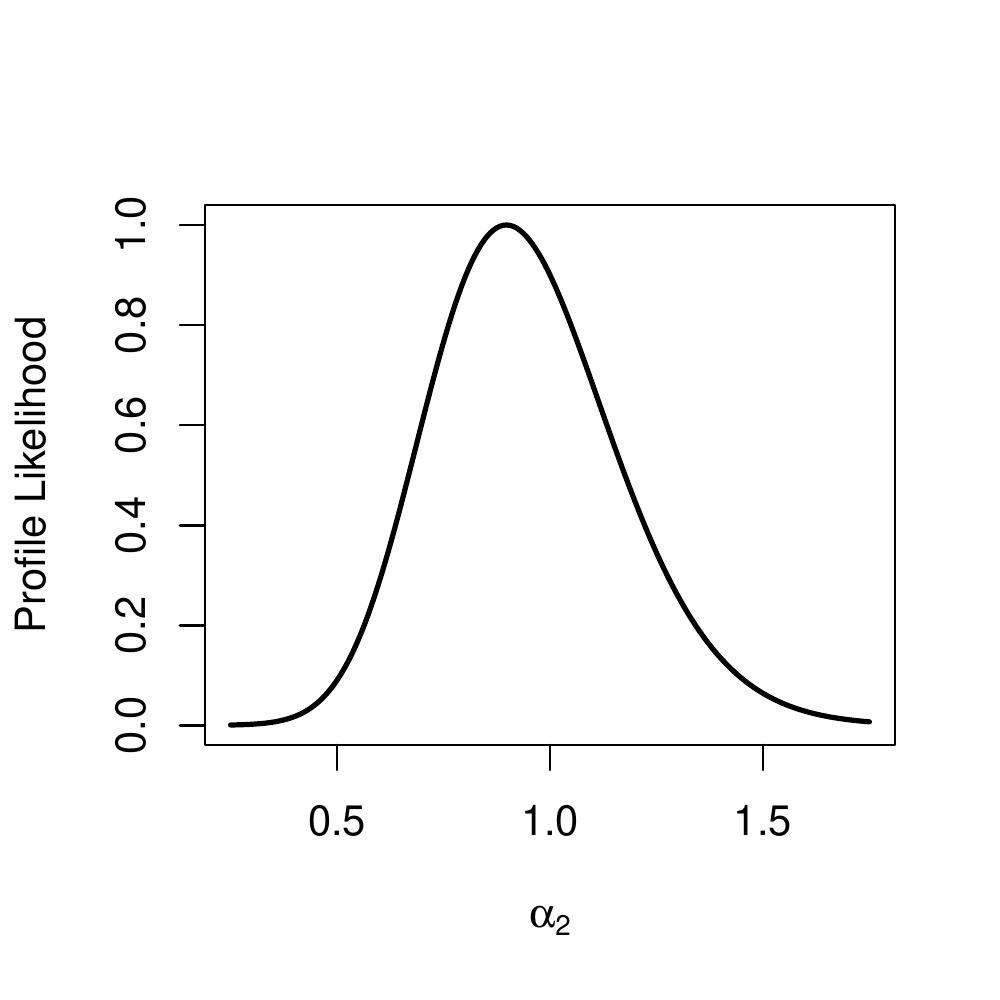} &
    \includegraphics[scale = 0.45]{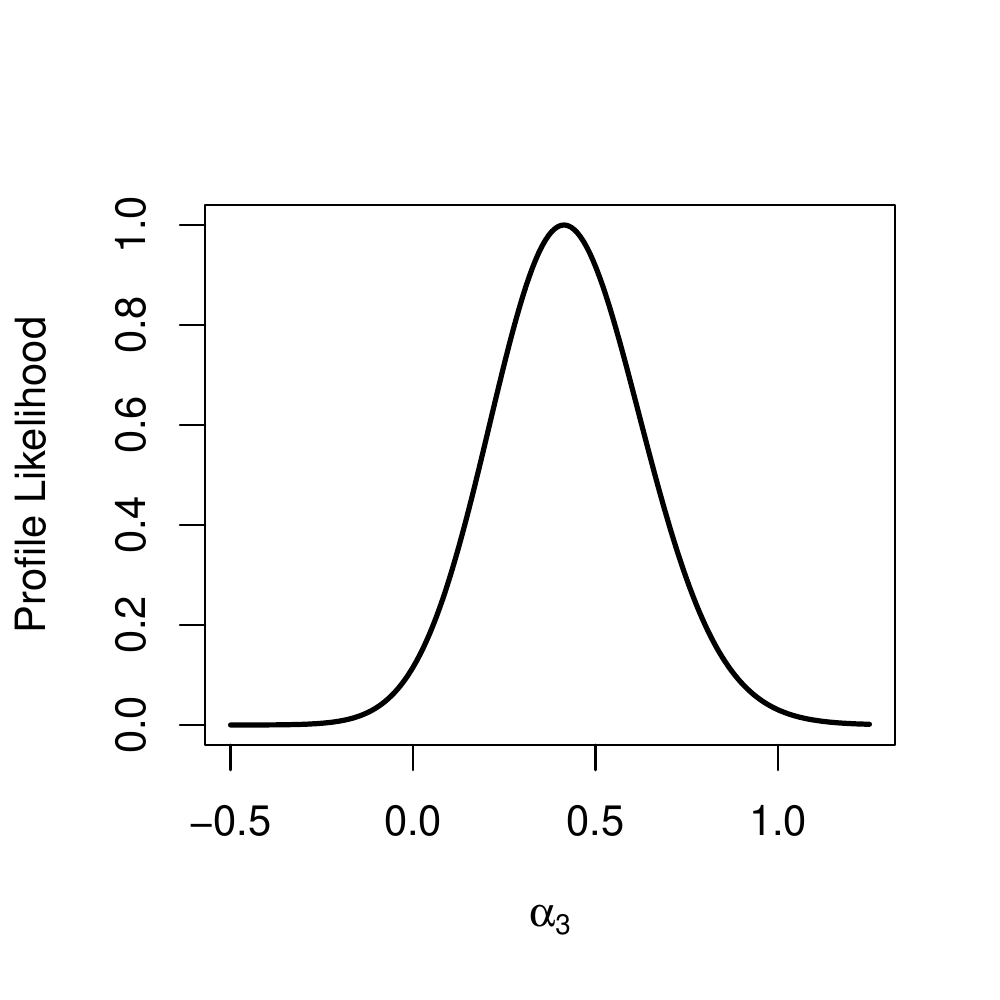} \\
    (d) & (e) & (f) \\
    \includegraphics[scale = 0.45]{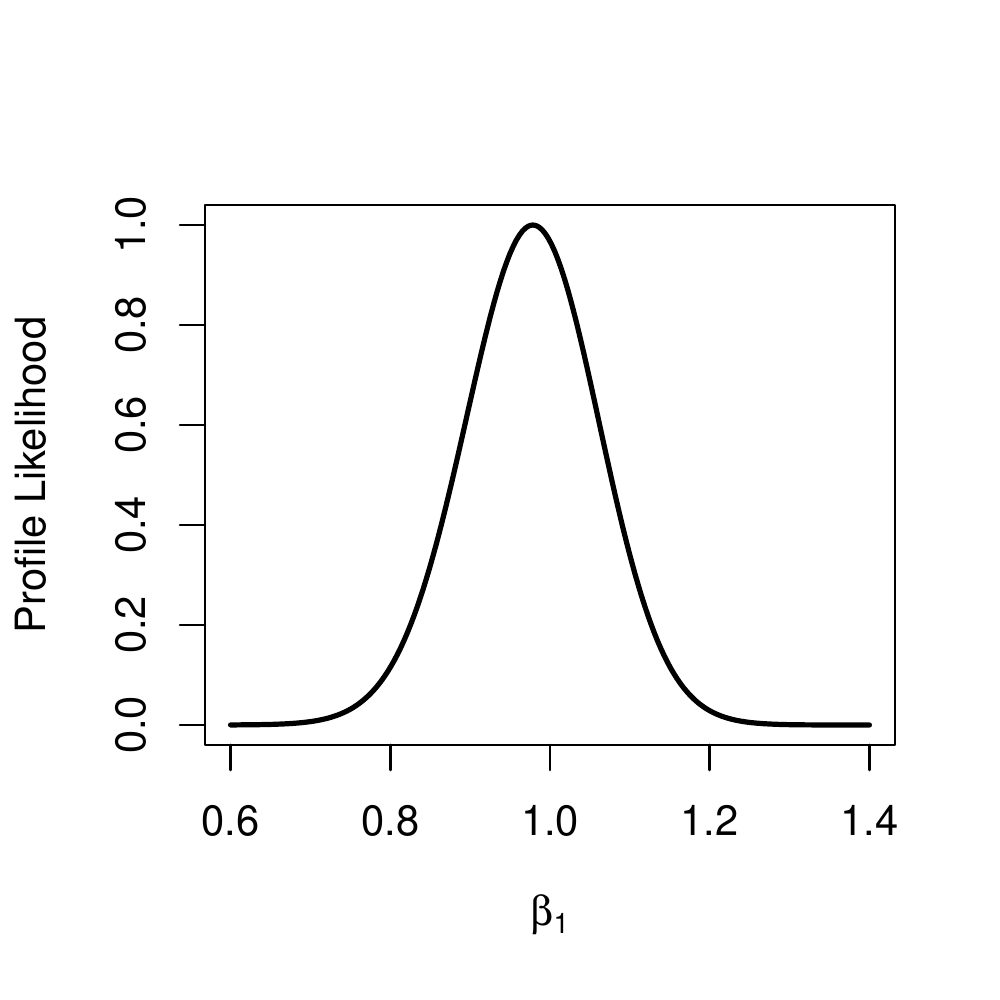} &
    \includegraphics[scale = 0.45]{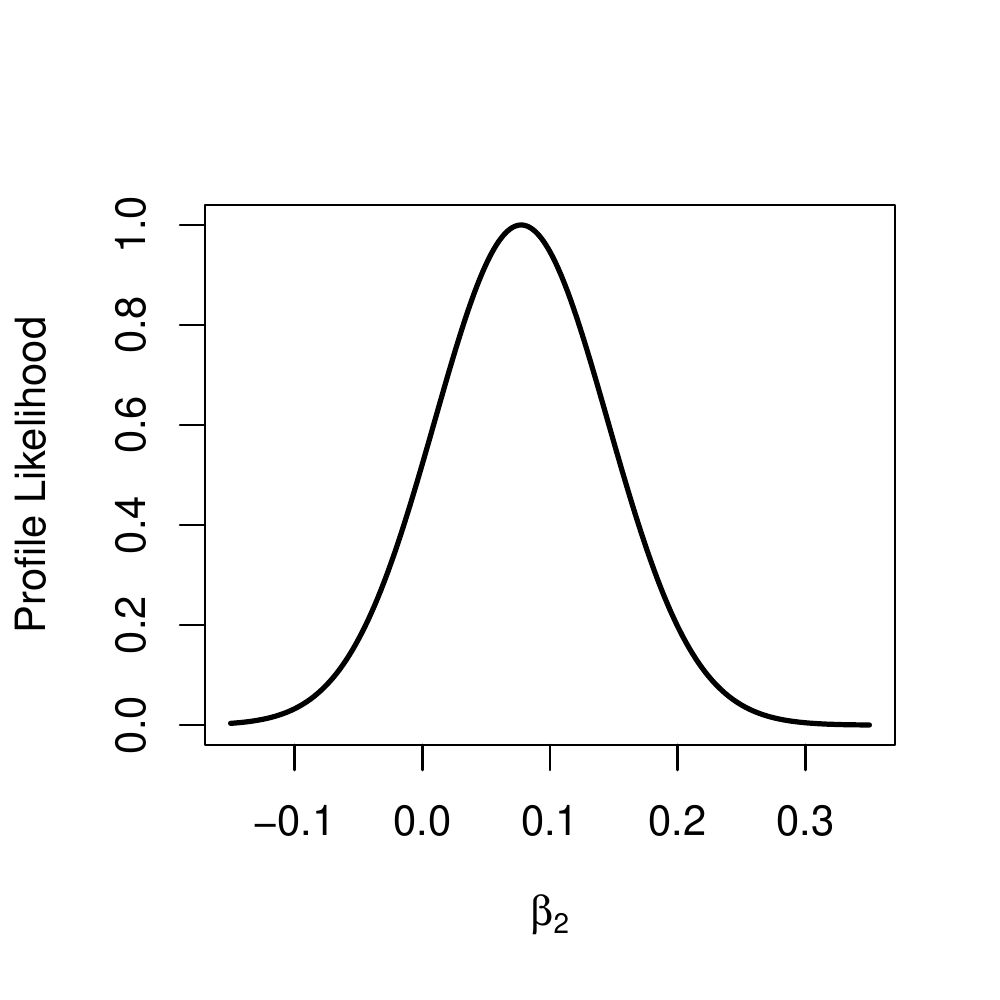} &
    \includegraphics[scale = 0.45]{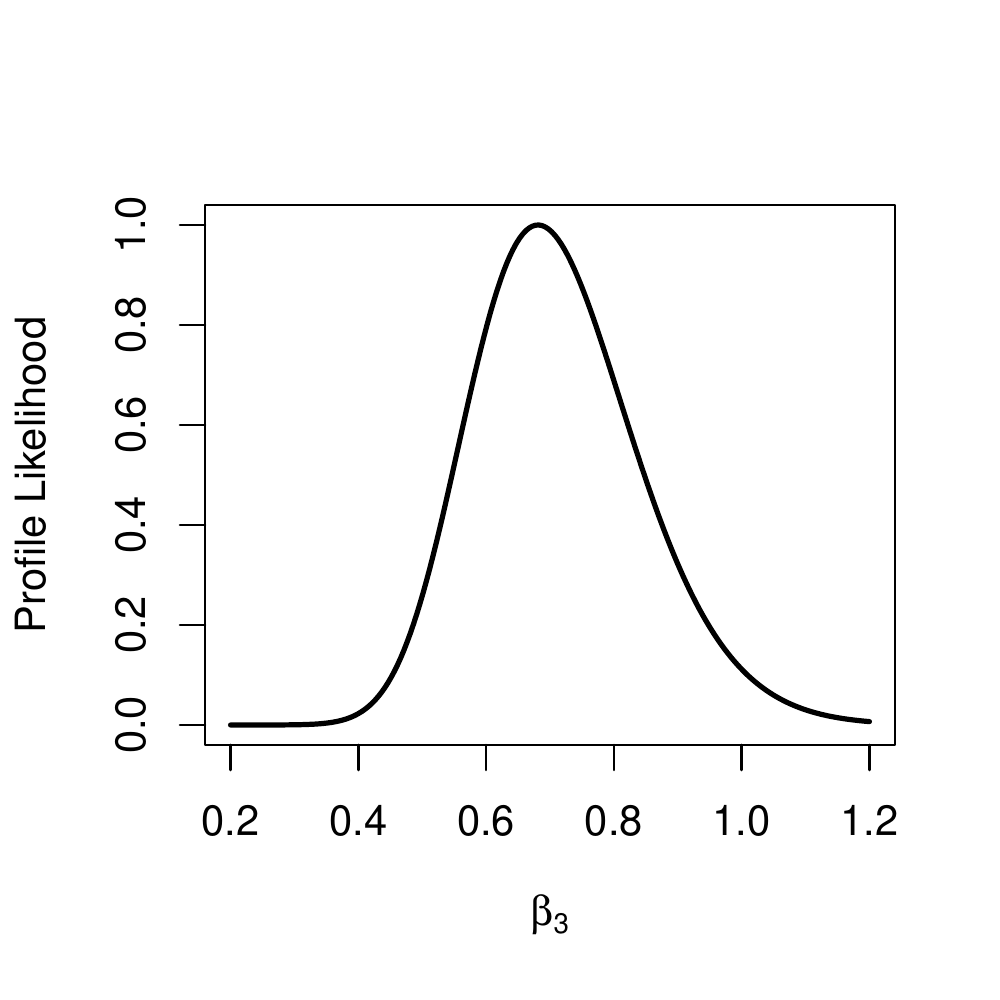} \\
    (g) & (h) & (i) \\
    \includegraphics[scale = 0.45]{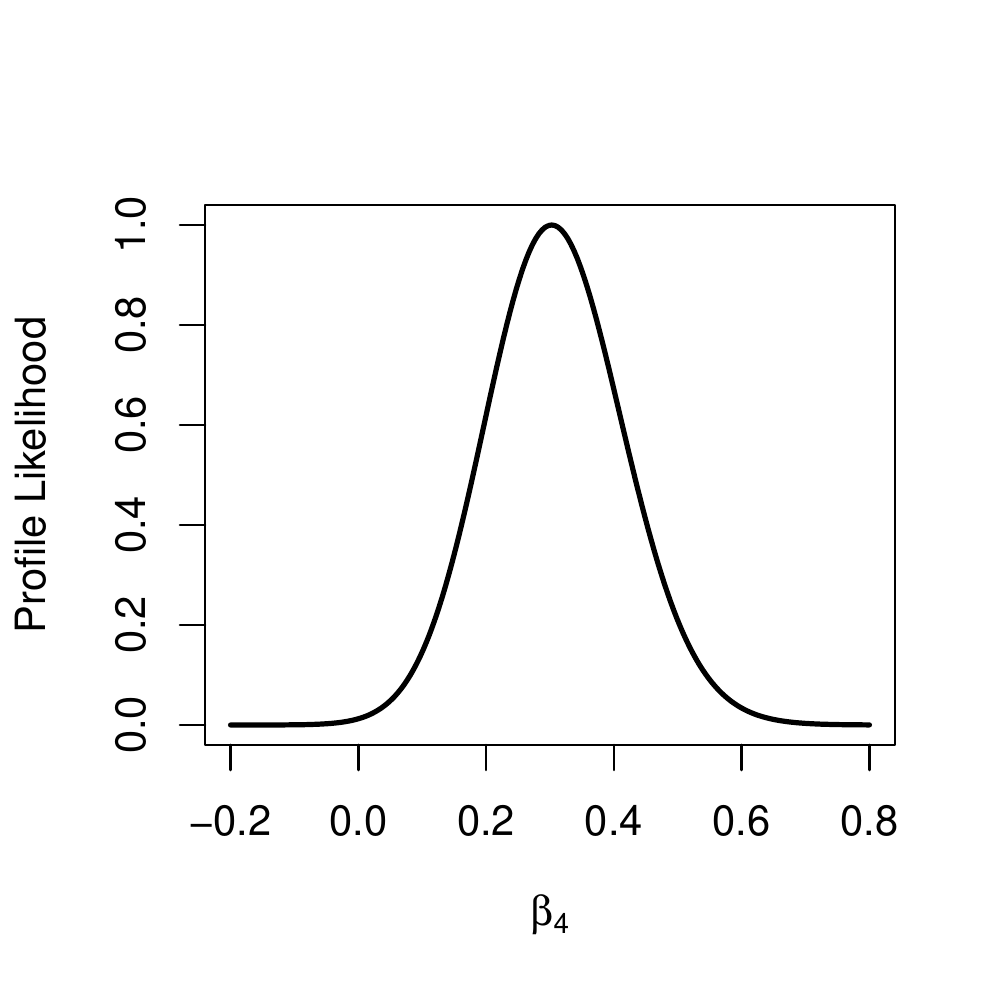} &
     &
     \\
    (j) &  &    
    \end{tabular}
    \caption{Leukemia cancer data. Profile likelihood plots for the parameters of the PGW-GH model.}
    \label{fig:Ex2-PGW-GH}
\end{figure}

We fit the GH model \eqref{eq:GH} with PGW baseline hazard (denoted as PGW-GH) with the time-level covariates \texttt{age}, \texttt{wbc} (white blood cell count at diagnosis) and \texttt{tpi} (Townsend score, which is a measure of deprivation), and hazard level covariates \texttt{age}, \texttt{sex}, \texttt{wbc} and \texttt{tpi}. Table \ref{tab:MLE2} shows the values of the maximum likelihood estimates of the parameters of this model. We first notice that the estimate of the parameter $\gamma$ is far from one but, in order to discard inferential issues, we need to inspect the different criteria.
In this line, Figure \ref{fig:Ex2-PGW-GH} shows the relative profile likelihood functions of the parameters of the PGW-GH model. Indeed, in all cases we notice that the profile likelihood functions are unimodal and seem to quickly decrease to zero. Thus, in this example we do not observe problems with multimodality or flat ridges of the profile likelihoods. Consequently, there is no evidence of practical non-identifiability for this model and this data set.
Now, the minimum KL divergence between the fitted PGW baseline hazard and the Weibull family is equal to $0.056$, while the upper bound in criterion \eqref{Ineq:KLD} is $0.0178$, and the effective sample size in $961$. Thus, the KL divergence criterion \eqref{Ineq:KLD}, with $M=0.05$, does not indicate near-redundancy of parameters of the PGW-GH model.
Also, the minimum Hellinger distance between the fitted PGW baseline hazard and the Weibull family is equal to $0.101$. Thus, the Hellinger distance criterion \eqref{Ineq:Hellinger} with $\kappa=0.05$ does not indicate near-redundancy of the parameters of the PGW-GH model. 
{The estimated probability (based on $B=1000$ bootstrap samples) that criteria \eqref{Ineq:KLD} and \eqref{Ineq:Hellinger} hold is zero}.
On the other hand, the smallest standardized eigenvalue of the Hessian Matrix of the log-likelihood function is $0.0035$. 
\color{black} Consequently, the Hessian method (with threshold $0.001$) does not indicate near-redundancy of parameters of this model and data. The bootstrap probability of the Hessian method with $0.001$ threshold is $0$. Thus, the Hessian method does not indicate near-redundancy of parameters of this model and data, even after accounting for uncertainty. \color{black}

For comparison, we now fit the GH model \eqref{eq:GH} with Exponentiated Weibull baseline hazard (EW-GH), as an alternative flexible model as well as simpler identifiable models such as the AFT model with PGW baseline hazard (PGW-AFT), PH model with PGW baseline hazard (PGW-PH), and the AFT model with Weibull baseline hazard (W-AFT). 
Table \ref{tab:MLE2} shows the MLEs and Akaike information criterion (AIC) for the different models. The AIC favors the EW-GH model overall, followed by the PGW-GH, respectively. This indicates that, although the different distributions can capture similar shapes, they are theoretically different and one of them may offer a better fit for a specific data set. 

\begin{table}[ht]
\begin{center}
\begin{tabular}{| c | c | c | c | c | c |  }
\hline
 & \multicolumn{5}{ c| }{Model} \\ \hline
Parameter & PGW-GH & PGW-PH & PGW-AFT &  EW-GH &  W-AFT\\ \hline
Scale $\widehat{\sigma}$ &0.095 & 0.139 & 0.093 &  0.006 & 1.152 \\ 
Shape $\widehat{\nu}$ & 1.006 & 0.815 & 1.014 &  0.219 & 0.575 \\ 
Shape $\widehat{\gamma}$ &  3.474 & 2.570 & 3.504 &  8.957  & -- \\ 
$\texttt{age}_t$  $\widehat{\alpha}_1$ & 0.911 & -- & -- & 0.886 & -- \\ 
$\texttt{wbc}_t$  $\widehat{\alpha}_2$ & 0.898 & -- & -- &  0.847 & -- \\ 
$\texttt{tpi}_t$  $\widehat{\alpha}_3$ & 0.413 & -- & -- &  0.407 & -- \\ 
\texttt{age} $\widehat{\beta}_1$ &  0.979 & 0.536 & 1.028 &  0.969 & 0.957\\
\texttt{sex} $\widehat{\beta}_2$ &  0.077 & 0.064 & 0.081 &  0.071 & 0.117 \\
\texttt{wbc} $\widehat{\beta}_3$ & 0.681 & 0.217 & 0.484 &  0.656 &0.371 \\
\texttt{tpi} $\widehat{\beta}_4$ & 0.303 & 0.097 & 0.216 &  0.302 & 0.159\\  
AIC &  1537.239 & 1586.483 & 1539.478 & \textbf{1534.681} & 1630.28\\ \hline
\end{tabular}
\caption{Leukemia data. Maximum likelihood estimation summary.}
\label{tab:MLE2}
\end{center}
\end{table}

\clearpage
\section{Discussion}\label{sec:discussion}

Parametric models have regained popularity in survival analysis thanks to the availability of richer formulations of survival regression models that allow for capturing complex effects, such as time-level, hazard-level, and nonlinear effects of the covariates (see \citealp{eletti:2022} for a general overview). However, an increase in model complexity also carries an inferential cost, which may be reflected, in mild scenarios, as wider confidence intervals or, in more extreme scenarios, in flat or multimodal likelihood functions. These types of inferential problems have been reported for the GH model \eqref{eq:GH}, and we have shown that such problems can be classified through the concepts of near-redundancy of parameters and practical non-identifiability. More specifically, we have shown that near-redundancy and practical non-identifiability problems appear in the GH model when the baseline hazard of the fitted model is close to the Weibull family.
\color{black} 

We have introduced a method for detecting near-redundancy in GH regression models based on the distance of the fitted model to a nested non-identifiable model. In addition, we have explored the use of the Hessian method to detect near-redundancy, while practical non-identifiability has been studied using of the profile likelihood function. We have also introduced the use of bootstrap methods to incorporate the uncertainty on the parameter estimation into the methods for detecting near-redundancy. In practice, detecting near-redundancy and practical non-identifiability is useful to understand the limitations in fitting a more complex model. For instance, the use of asymptotic results (such as standard errors and normal confidence intervals) may not be valid or may produce inaccurate results in the presence of these problems. Moreover, the presence of these inferential problems may put into question the use of such complex models for important tasks such as prediction, policy or decision making.

The simulation study shows that the distance-based methods and the Hessian method have comparable performance in detecting near-redundancy. Moreover, comparing the cases classified as near-redundant with those cases classified as practically non-identifiable reveals a strong connection between the presence of these two inferential problems in the GH model. 
The simulation results also show that the interplay of model complexity, sample size, and censoring rate plays an important role in the appearance of near-redundancy and practical non-identifiability in the GH model. In problematic samples in our simulation study and case studies (\textit{i.e.}~leading to near-redundancy and/or practical non-identifiability), we noticed that the profile likelihood may not only contain flat ridges, but also local maxima and abrupt changes in the derivative of the profile likelihood.

Whenever a case with near-redundancy or practical non-identifiability is obtained, a possible alternative is to use a simpler identifiable model that can be written in terms of a smaller set of parameters (\textit{i.e.}~not containing a parameter-redundant model). These simpler models include the PH and AFT models. Indeed, recent references \citep{simpson:2022} have suggested the use of model selection aided by information about practical non-identifiability of the models under comparison, and that more complex models should be used when there is enough data to estimate the model parameters. However, there are other possible solutions, such as increasing the effective sample size (either by adding new samples or increasing the follow-up time). In summary, the three possible solutions are: reducing censoring, increasing the sample size, or using a simpler identifiable model. We also recommend looking for other problems in the likelihood function, beyond flat ridges. Indeed, the presence of flat ridges also depends on the parameterization, as one parameterization may lead to flat ridges while another parameterization of the same model may lead to a different behavior.
\color{black} 

Although we have focused on the study of right-censoring (the most common type of censoring), our conclusions are likely to apply to samples with left-censoring, interval-censoring, and truncation. Other potential extensions of our work consist of analyzing near-redundancy and practical non-identifiability in other types of flexible parametric survival models, including generalized additive models \citep{eletti:2022}, other general classes of survival models \citep{muse:2022}, cure models \citep{hanin:2014}, or distributional-regression models \citep{rigby:2005,burke:2020}.  The fact that simpler models represent an alternative to avoid problematic inferential cases, points to the need for developing formal model selection tools in the context of the GH model (see \citealp{rossell:2019} for a general overview on model and variable selection in survival models). \color{black} Finally, other areas in Statistics where flat likelihoods appear, such as inference in circular models \citep{miyata:2022,johnson:2022}, may also benefit from linking inferential problems in those models with the concepts of near-redundancy and practical non-identifiability. \color{black}

\section*{Acknowledgements}
We thank Prof.~Yannick Baraud for his guidance on the works by Le Cam that led to the development of criterion \eqref{Ineq:Hellinger} based on the Hellinger distance.
We also thank two reviewers for the constructive comments provided.

\bibliographystyle{plainnat}
\bibliography{references}  

\end{document}